\documentclass[english,
]{aa}
\usepackage{epsf,amsfonts,amssymb,graphicx,fancyheadings,caption}
\usepackage{natbib}
\usepackage{babel}
\usepackage{txfonts}
\usepackage{rotating}
\usepackage{url}
\usepackage{float}
\usepackage{enumitem}

\bibpunct{(}{)}{;}{a}{}{,}   

\begin{document}

\title{The dearth of dustless HdC stars in the Magellanic Clouds: A metallicity-driven white dwarf merger mass effect}

\author{
P.~Tisserand\inst{1},
A.~J.~Ruiter\inst{2},
E.~Cheng\inst{3},
J.~Soon\inst{4},
A.~M\"oller\inst{5},
C.~L.~Crawford\inst{6},
G.~Bowden\inst{3},
I.~R.~Seitenzahl\inst{4},
S.~J.~Murphy\inst{7},
C.~A.~Onken\inst{4},
C.~Wolf\inst{4},
G.~C.~Clayton\inst{8,9}
}

\institute{
Sorbonne Universit\'e, CNRS, UMR 7095, Institut d'Astrophysique de Paris, 98 bis bd Arago, 75014 Paris, France  \and
Mathematical Sciences Institute, The Australian National University, ACT 2601, Australia \and
School of Engineering and Technology, University of New South Wales, Canberra, ACT 2600, Australia \and
Research School of Astronomy and Astrophysics, Australian National University, Cotter Rd, Weston Creek ACT 2611, Australia \and
Centre for Astrophysics and Supercomputing, Swinburne University of Technology, John St, Hawthorn, VIC 3122, Australia \and
Sydney Institute for Astronomy (SIfA), School of Physics, University of Sydney, NSW 2006, Australia \and
School of Science, University of New South Wales, Canberra, ACT 2600, Australia \and
Department of Physics \& Astronomy, Louisiana State University, Baton Rouge, LA 70803, USA \and
Space Science Institute, 4765 Walnut St., Suite B Boulder, CO 80301, USA
}

\offprints{Patrick Tisserand; \email{tisserand@iap.fr}}

\date{}

\abstract 
{Hydrogen-deficient carbon (HdC) stars are rare supergiants believed to form through the merger of two white dwarfs (WDs). They comprise the dust-producing R Coronae Borealis (RCB) stars and the dustless HdC (dLHdC) stars. While dedicated Galactic surveys find the two subclasses in roughly equal numbers, no dLHdC star had previously been identified in the Magellanic Clouds, despite 30 known RCB stars in these two galaxies.}
{We present the first systematic search for dLHdC stars in the Large and Small Magellanic Clouds, aiming to measure the dLHdC-to-RCB ratio in environments of distinct metallicities and well-constrained star formation histories.}
{Candidates were selected from the Gaia DR3 Magellanic catalogue using colour--magnitude cuts combining Gaia, SkyMapper DR4, and VMC DR6 photometry, complemented by a random-forest classifier trained on known Galactic dLHdC stars. The search reached an overall detection efficiency of $\sim$96\%. All candidates were followed up spectroscopically with WiFeS on the ANU 2.3~m telescope.}
{We discovered the first two Magellanic dLHdC stars: SP77 29-5 in the LMC and SMSS 005551.43-723139.2 in the SMC, the latter belonging to the rare Sr-rich subclass. The resulting dLHdC-to-RCB number ratio of $\sim$1:15 overall is far below the near-unity Galactic value.}
{We argue this deficit is intrinsic rather than observational. Convolving  the delay time distribution from a sub-class of white dwarf mergers with each Cloud's star formation history, we estimate that $\sim$80\% (LMC) and about two-thirds (SMC) of HdC stars originate from mergers with short delay times, which at low metallicity are biased towards higher total WD masses, and thus towards the luminous regime associated with RCB stars rather than the fainter dLHdC population.}

\keywords{methods: data analysis -- stars: carbon -- stars: chemically peculiar -- supergiants -- binaries: close -- Magellanic Clouds}

\authorrunning{Tisserand, P. }
\titlerunning{New dLHdC stars in the Magellanic Clouds}

\maketitle

\section{Introduction \label{sec_intro}}

Hydrogen-deficient carbon (HdC) stars are a rare class of supergiant stars with atmospheres composed almost entirely of helium ($\sim$98\%, $\sim$10$^4$ times more abundant than hydrogen) and about 1\% carbon. They span a broad range of absolute magnitudes (M$_V$ $\sim$ $-5$ to $-1.5$~mag) and effective temperatures \citep{Tisserand2022}. A temperature-based classification, from HdC0 ($\sim$8750~K) to HdC7 ($\sim$4000~K), was defined by \citet{Crawford2023}. 

HdC stars are divided into two main subgroups. The first is composed of the dust-producing R~Coronae~Borealis (RCB) stars, known for their sudden photometric declines due to carbon dust cloud formation along the line of sight \citep{Clayton1996Review,Woitke1996}. The second is composed of the dustless HdC (dLHdC) stars, which show little or no dust formation \citep{Tisserand2022}. Compared to RCBs, dLHdC stars are typically warmer (5000~$< T_{\rm eff} <$~8000~K) and about two magnitudes fainter within that range. At even higher temperatures ($T_{\rm eff} \geq 10{,}000$~K), the extreme helium (EHe) stars are believed to be closely related objects, possibly representing a later contraction phase of HdC stars \citep{Jeffery2008_EHeReview}.

HdC stars are increasingly believed to represent the giant phase following the merger of two white dwarfs (WDs) --- typically a carbon-oxygen WD and a helium-rich WD --- that is expected to last between a few 10$^4$ and a few 10$^5$ years  \citep{Webbink1984, Jeffery2011, Crawford2024}. Their observed luminosity range is consistent both with the population synthesis models, which predict a distribution of total WD masses and mass ratios at the end of close binary evolution \citep{Tisserand2020, Tisserand2022} and with the brightness of the giant phase calculated by WD merger models \citep{Crawford2024}. Thus, the dLHdC stars, being of lower luminosity, could be used to probe the low-mass regime of He+CO WD systems. Within this framework, the origin of the observed diversity in effective temperatures remains an open question, although metallicity may play a key role. The wide range of expected merger delay times --- the interval between the
formation of the original binary system and the final white dwarf merger --- from about 250 million years up to a Hubble time, together with the detection of HdC stars across all Galactic substructures
\citep[][see their Fig. 19]{Tisserand2024a}, strongly supports the presence of a broad range of metallicities among HdC atmospheres.

Abundance analyses further support a merger origin: the chemical compositions of HdC atmospheres retain signatures of the complex evolutionary histories of their progenitor binary systems, involving phases such as post-AGB evolution, common-envelope interaction, nova-like activity, and the final WD coalescence. Indeed, HdC stars display remarkably high abundances of $^{18}$O, with $^{16}$O/$^{18}$O ratios more than 500 times lower than in the Sun \citep{Clayton2007, Garcia-Hernandez2009}, and for dLHdC stars, even below unity \citep[$^{16}$O/$^{18}$O~$<$~1;][]{Karambelkar2022, Mehla2025}. They also exhibit enhanced fluorine abundances \citep{Pandey2008}, overabundances of $s$-process elements \citep{Jeffery2011, Crawford2022}, and, in some cases, high lithium levels \citep{Asplund2000}.

In that context, HdC stars represent a rare observational window into the endpoint of binary stellar evolution, where the merger of two white dwarfs leads not to a thermonuclear explosion but to the formation of a stable, hydrogen-deficient supergiant. Understanding their physical properties and frequency therefore provides crucial empirical constraints on the outcomes of white-dwarf mergers---a process also relevant to the progenitors of Type~Ia supernovae  \citep{2025A&ARv..33....1R} and the sources of low-frequency gravitational waves detectable by \textit{LISA} \citep{Kremer2026__LISA_RCB}.

A competing formation scenario is the final helium-shell flash model, in which a star on the verge of becoming a white dwarf undergoes a late helium-shell ignition and expands back to supergiant dimensions \citep{Renzini1990, Iben1996}. Although this scenario appears less consistent with most observed HdC properties, it cannot be entirely excluded; a fraction of the nearly 200 HdC stars known to date may have formed through this channel \citep{Clayton2012Review}.

The number of known RCB stars has increased dramatically over the past two decades, thanks to extensive microlensing monitoring campaigns \citep{Alcock2001, Zaniewski2005, Tisserand2008, Tisserand2009, Tisserand2013} and the all-sky infrared surveys conducted by 2MASS and WISE \citep{Tisserand2012, Tisserand2020, Karambelkar2024}. Until very recently, however, only four bright Galactic dLHdC stars were known (HD~137613, HD~148839, HD~173409, and HD~182040) plus HE 1015-2050 identified in the Galactic halo by \citet{Goswami2010}. The situation changed with the advent of the Gaia~DR3 survey, whose astrometric and photometric data led to the identification of 27 additional dLHdC stars in the Milky Way \citep{Tisserand2022}. Although some of these may be classical RCB stars temporarily in a phase of low dust production, the majority appear to be genuine dLHdC stars. Adding further complexity, recent studies suggest that the dust-production rate of RCB stars decreases with increasing effective temperature \citep{Tisserand2024b,Crawford2025}.

The Magellanic Clouds provide an ideal environment in which to extend our study of HdC stars. Their well-constrained distances, low foreground extinction, and distinct metallicities offer a unique opportunity to explore how chemical composition and stellar population context influence HdC formation and evolution. Following the discovery of the first three Magellanic RCB stars -- W~Mensae \citep{Luyten1927, Feast1956, Rodgers1970}, HV~5637, and HV~12842 \citep{Payne-Gaposchkin1971, Feast1972} -- subsequent surveys have brought the total number of Magellanic RCB stars to about 30 \citep{Alcock1996, Alcock2001, Tisserand2004, Tisserand2009, Tisserand2020}. However, no dLHdC star has yet been reported in either of the Magellanic Clouds.

In this paper, we present our systematic search for dLHdC stars in the Magellanic Clouds. We describe our selection methodology based on two analyses, a classical selection with predefined colour and brightness cuts and a subsequent one that used a machine learning random-forest algorithm (Sect.~\ref{sec:ana}), then we describe the newly found Magellanic dLHdC stars confirmed after a spectroscopic follow-up in Sect.~\ref{sec:New-dLHdC}. We discuss in Sect.~\ref{sec:discus} the implications of our findings for the merger-formation paradigm of HdC stars in light of the specific Magellanic stellar population. A summary of our search is given in Sect.~\ref{sec:concl}. 

\section{Analysis \label{sec:ana}}

In this study, we used as reference catalogue the \textit{Gaia} Magellanic-star sample published by \citet{Luri2021} where a filter to remove a large fraction of foreground Galactic stars was applied. It covers large sky areas corresponding to circular regions of 20$^\circ$ and 11$^\circ$ radii centred on the LMC and SMC, respectively. Consequently, the Magellanic Bridge is included in the analysis, and a small number of stars are common to both regional catalogues. Specifically, we focused our analysis on one stellar subsample: the blue-loop (hereafter BL) population. This subsample was defined as a polygonal selection in the $G$ versus $(G_{\mathrm{BP}} - G_{\mathrm{RP}})$ colour--magnitude diagram. It is of interest to us as it overlaps the locus of the known Galactic dLHdC stars (see Fig.~\ref{fig:HRFig}). The BL subsample contains 261,929 (86,212) stars in the LMC (respectively SMC). 

The reference BL Gaia catalogue was cross-matched with other optical and IR catalogues using a 0.5" radius with the CDS X-Match service and the TOPCAT software cross-matching algorithm.

Infrared photometry was primarily obtained from the VISTA Magellanic Cloud (VMC) survey, DR6 release \citep{Cioni2011, VMC_DR6_Cioni2023}, providing high-quality $J$- and $K_s$-band measurements. We adopted the recommended aperture photometry (``aper3'') measured with a 1" radius, which offers excellent photometric precision. Within the VMC DR6 footprint, 98\% of the Gaia BL stars were associated with a VMC star (about 86\% of all Gaia BL stars). The missing stars are either located in the most crowded Magellanic regions or in a $\sim$3\,deg$^2$ area centred on RA$\sim$78.8$^\circ$ Dec$\sim$-72.2$^\circ$ where the $J_\mathrm{aper3}$ magnitudes are missing. For completeness, we supplemented the VMC IR data with the 2MASS-6X catalogue \citep{2MASS_6X_Cutri2012}, which reaches sensitivities $\sim$1\,mag deeper than the all-sky 2MASS survey in crowded regions. The 2MASS-6X photometry was used only when VMC DR6 data were unavailable.

We further used optical PSF photometry from the SkyMapper DR4 survey \citep{SkyMapperDR4-Onken2024}, which provides valuable coverage of the blue spectral region through its narrow u and v filters. The principal issue comes from the difference in resolution and magnitude limits between the SkyMapper blue survey and the Gaia+VMC surveys. As presented on the HR diagram in Fig.~\ref{fig:HRFig}, only the brightest and bluest BL stars were catalogued, i.e., about 27200 stars or $\sim$12\% of the VMC DR6 BL stars. Fortunately, it covers well the luminosity and colour ranges of the known Galactic dLHdC stars.

Interstellar reddening corrections were applied using the extinction maps of \citet{Skowron2021}, based on red-clump stars monitored by the OGLE-IV survey. These maps provide a resolution of 1.7'~$\times$~1.7' in the central regions of the Magellanic Clouds. In the outskirts, when such reddening values were missing, or the box size of the \citet{Skowron2021} map exceeded 7 arcmin, the interstellar reddening values of \citet{Schlafly-Finkbeiner2011-Reddening} were used instead. The extinction coefficient R$_\lambda$ = A$_\lambda$ / E(B-V) for a broadband filter is conventionally tabulated using a representative source spectral energy distribution (SED) with a hydrogen-dominated continuum \citep{1989ApJ...345..245C, 1999PASP..111...63F, Schlafly-Finkbeiner2011-Reddening}. The SEDs of HdC stars differ markedly from those of hydrogen-rich stars, both in their continuum opacity sources and in the strong C$_2$ and CN molecular bands that redistribute the emergent flux across the optical and near-IR \citep{Asplund1997,Asplund2000,Hema2012}. Because R$_\lambda$ depends on the filter-integrated SED, we computed a dedicated R$_\lambda$ grid from synthetic HdC spectra made from Hydrogen-deficient MARCS (Model Atmospheres in Radiative and Convective Scheme) atmospheric models \citep{1975A&A....42..407G,1976A&AS...23...37B,2008PhST..133a4003P,2008A&A...486..951G} for different effective temperatures ($T_\mathrm{eff}$ between 4000 - 7500 K). Thus, for all Magellanic stars studied, we used the following median extinction coefficient R$_b$ values for all the photometric bands \textit{b} used in this analysis: 4.92, 4.68, 3.39 for SkyMapper u, v, g;  2.57, 3.22, 1.97 for Gaia G, G$_{BP}$, G$_{RP}$, and 0.90, 0.37 for VMC J, K$_S$ and 2MASS J and K bands. Synthetic photometry was computed from the model spectra using the pyphot package \citep{pyphot}, with the transmission curves of all passbands used in this analysis retrieved from the SVO Filter Profile Service \citep{2012_SVO,2020_SVO}.



We applied colour and magnitude selection criteria to search for Magellanic dLHdC candidates, complemented by a machine-learning classification using a random-forest algorithm. In both analyses, known Galactic dLHdC stars were used as references. The details of the analysis are presented in the following sections.

\subsection{Spectroscopic follow-up}

All candidates selected from both analyses were followed up spectroscopically with the Wide Field Spectrograph (WiFeS) instrument \citep{2007Ap&SS.310..255D} attached to the automated 2.3m telescope \citep{Price2024-Automatic2.3m} of the Australian National University at Siding Spring Observatory (SSO). WiFeS is an integral-field spectrograph permanently mounted at the Nasmyth A focus. It provides a 25"$\times$38" field of view with 0.5" sampling along each of the twenty-five 38"$\times$1" slitlets. The visible wavelength interval is divided by a dichroic at around 560 nm, feeding two essentially similar spectrographs. We obtained spectra with two-pixel resolution of 2~\AA\ (R$\sim$3000) and wide wavelength coverage, from 340 to 960 nm. It was reduced using the PyWiFeS data reduction pipeline\footnote{The pipeline was updated with a version 2.1, see \url{https://github.com/PyWiFeS/pywifes} and an upcoming publication.} \citep{2014Ap&SS.349..617C}.

\subsection{Known Galactic dLHdC stars}

Of the 32 known Galactic dLHdC stars listed in \citet[Table 2]{Tisserand2022}, only two, HE 1015-2050 \citep{Goswami2010} and A166 \citep{Tisserand2022}, which are both expected to be members of the Galactic halo \citep{Tisserand2024a}, were not kept in our analysis due to a lack of distance estimate. Gaia DR3 does not provide a parallax for HE 1015-2050 and A166's Gaia astrometric quality indicator, the RUWE (Renormalized Unit Weight Error), was too high \citep{2021A&A...649A...2L}.

We used photometric measurements in the following passbands: Gaia G, G$_{BP}$, G$_{RP}$, SkyMapper DR4 u, v, g, r, i, z, and 2MASS J and K. Only two known dLHdC stars, C542 and A814, did not have valid photometric measurements in u and v. The bright dLHdC stars, HD 137613, HD 148839 and HD 182040, did not have valid SkyMapper g measurements due to saturation.

As a distance proxy, we used the geometric distances and associated errors calculated by \citet{Bailer-Jones2021} and their respective extinction corrections were obtained from the 3D dust map created by \citet{Green2019} if available, otherwise, we relied on E(B-V) reddening values provided by \citet{Schlafly-Finkbeiner2011-Reddening}. The extinction coefficients R$_\lambda$ were calculated for each known dLHdC individually in all bands, since their values depend not only on their respective effective temperature but also on the level of interstellar reddening E(B-V), which reaches values near unity for some of them. 

\subsection{Method 1: Colour and luminosity selection cuts}

The Gaia BL stellar subsample was used to search for Magellanic dLHdC stars. We focused our analysis on stars presenting a valid IR counterpart and whose reddening E(B-V) parameter value was lower than 0.2 mag to avoid regions of star formation ($\sim$91\% and $\sim$94\% of the original stars sample in the LMC and SMC, respectively). We describe hereafter the series of selection cuts applied to identify candidates that were subsequently followed up spectroscopically. The selection cuts were defined using the known Galactic dLHdC stars as references. The effect of each selection cut is given in Table~\ref{tab:selectioncuts}. 

\begin{figure}[h]
    \centering
    \includegraphics[width=1.0\linewidth]{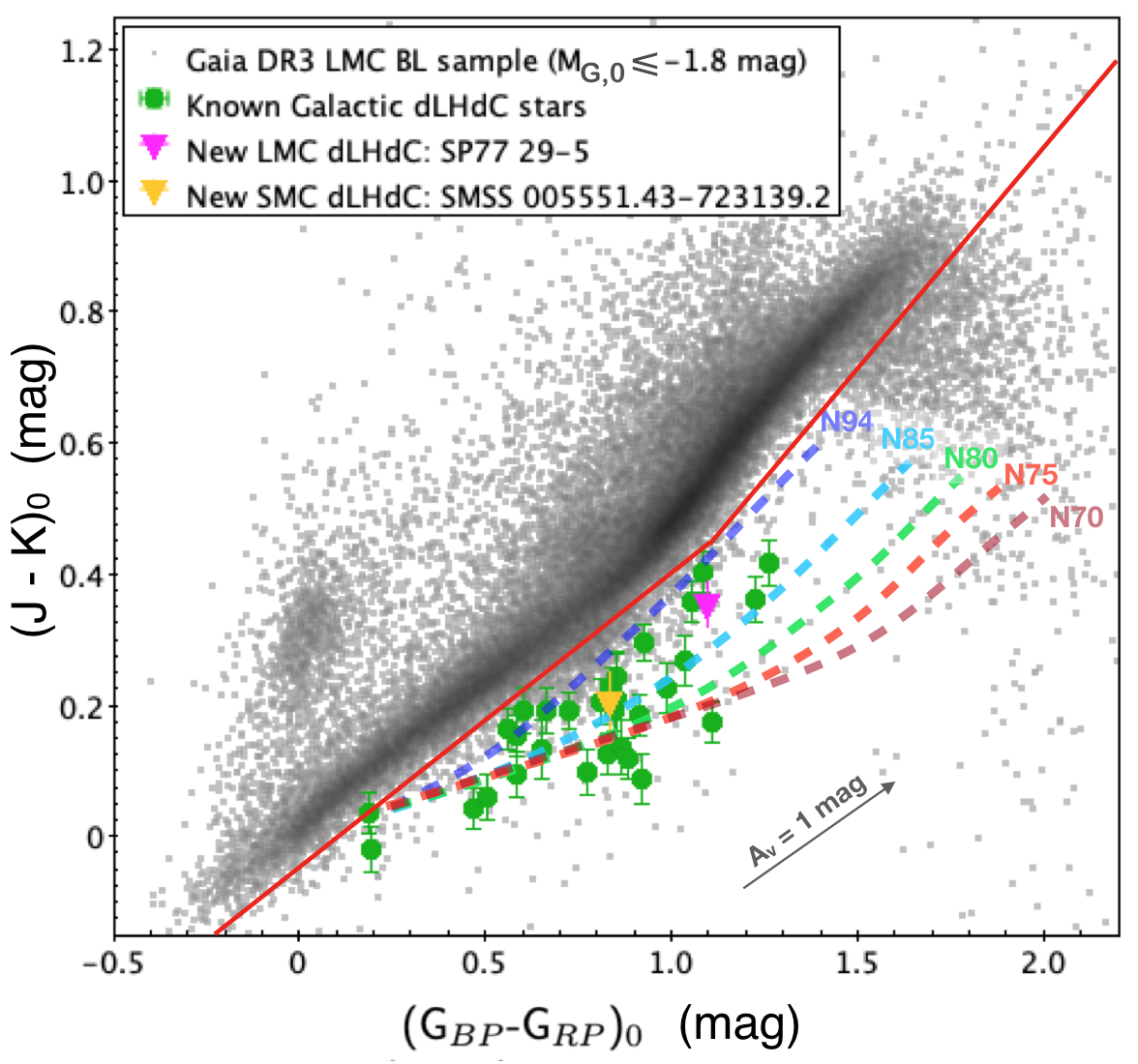}
    \caption{Diagram (J-K$_s$)$_0$ vs (G$_{BP}$-G$_{RP}$)$_0$. The LMC BL stars brighter than M$_{G,0}\leq$-1.8 mag are represented with grey points. The large green dots correspond to the known Galactic dLHdC stars while the newly discovered Magellanic dLHdC stars are represented with down triangles. The coloured dashed lines indicate the expected position of dLHdC stars based on synthetic spectra made from H-deficient MARCS models for different N abundances in the 4000 - 7500 K effective temperature range. The reddening vector corresponding to an extinction A$_V$=1 mag is presented with a black arrow. The selection cut \#1 is illustrated with red lines.}
    \label{fig:JKvsBpRp}
\end{figure}

\begin{figure*}[htbp]
    \centering
    \includegraphics[width=1.0\linewidth]{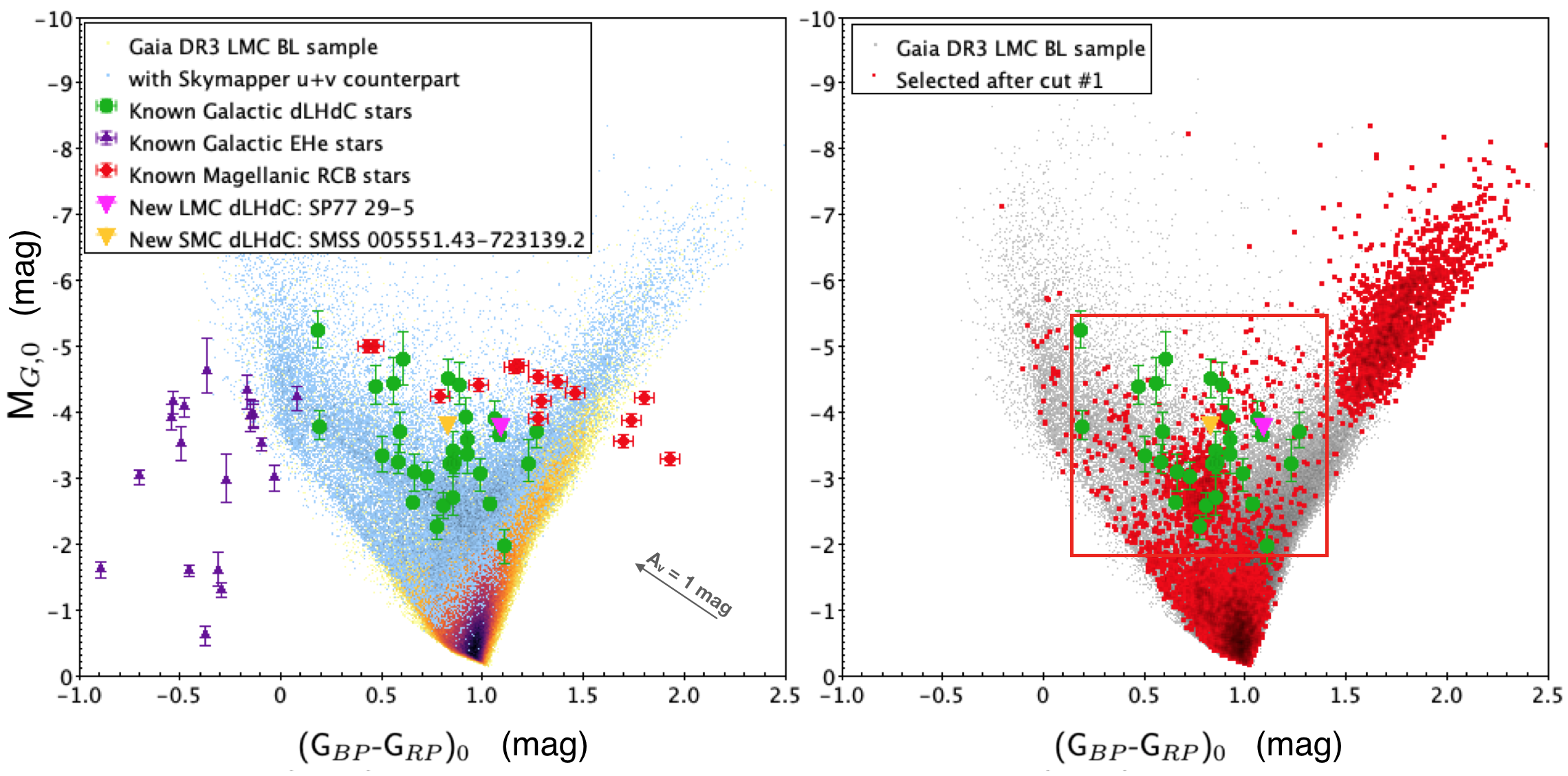}
    \caption{M$_{G,0}$ vs (G$_{BP}$-G$_{RP}$)$_0$ HR diagrams. In both diagrams, the green dots indicate the position of the known Galactic dLHdC stars, and the down triangles show those of the newly discovered Magellanic dLHdC stars. Left: distribution in yellow density scale of all Gaia DR3 LMC BL stars (the majority of them are fainter than M$_{G,0}>$-2 mag). The ones having valid SkyMapper DR4 u and v PSF photometric measurements are indicated in light blue. The position of the known Galactic EHe stars are in dark purple and some known Magellanic RCB stars for which we have photometric information in their quiescent state are shown in red. The reddening vector corresponding to an extinction A$_V$=1 mag is presented with a black arrow. Right: distribution in red of the selected BL stars after cuts \#1, on top of all Gaia DR3 LMC BL stars presented with grey dots. The red rectangle corresponds to the selection area applied in colour and luminosity with cut \#2.}
    \label{fig:HRFig}
\end{figure*}

Our first selection was applied in the (J-K$_s$)$_0$ versus (G$_{BP}$-G$_{RP}$)$_0$ colour-colour diagram (see Fig.~\ref{fig:JKvsBpRp}), thus using the two deepest surveys, which also presented the best photometric accuracy. About 98\% of the BL stars were removed at that initial stage. The known dLHdC stars stand out from the main locus of hydrogen-rich BL stars due to a difference in shape of the overall SED. For a similar G$_{BP}$-G$_{RP}$ colour, dLHdC stars appear bluer in the IR. Furthermore, using synthetic spectra made from Hydrogen-deficient MARCS atmospheric models for different nitrogen abundances ($[N]$ from 7.0 to 9.4) and effective temperatures ($T_\mathrm{eff}$ between 4000 - 7500 K), we can explain the spread observed in the known dLHdC stars distribution. The depth of the molecular CN spectroscopic features changes the J-K colour. The selection cuts are as follows:

\begin{equation}
\begin{array}{l}
    if \hspace{0.1cm}(G_{BP}-G_{RP})_0\leq1.1 : \\
     \hspace{1cm} (J-Ks)_0 \leq 0.444 \times (G_{BP}-G_{RP})_0 - 0.044
\end{array}
        \label{eq_JK_BPRP_a}
\end{equation}

\begin{equation}
\begin{array}{l}
    if \hspace{0.1cm}(G_{BP}-G_{RP})_0>1.1 : \\
     \hspace{1cm} (J-Ks)_0 \leq 0.680 \times (G_{BP}-G_{RP})_0 - 0.30
\end{array}
        \label{eq_JK_BPRP_b}
\end{equation}

Second, the remaining Gaia Magellanic BL stars were selected based on their luminosity (-5.5$\leq M_G \leq$-1.8 mag - that faint limit was replaced by the more stringent value of -2.0 mag if the IR information was provided by 2MASS-6X due to its poorer photometric resolution) and colour (0.15$\leq (G_{BP}-G_{RP})_0\leq$1.4 mag). The selected area is represented in Fig.~\ref{fig:HRFig}, right. It was designed to reject mostly the numerous faintest BL stars as well as the red supergiant stars. Distance moduli of 18.5 and 19.0 mag were used, respectively, for the LMC and SMC. Only 1338 stars ($\sim$0.4\%) of all Magellanic BL stars were selected at this stage (see Table~\ref{tab:selectioncuts}).

\begin{figure*}[htbp]
    \centering
    \includegraphics[width=1.0\linewidth]{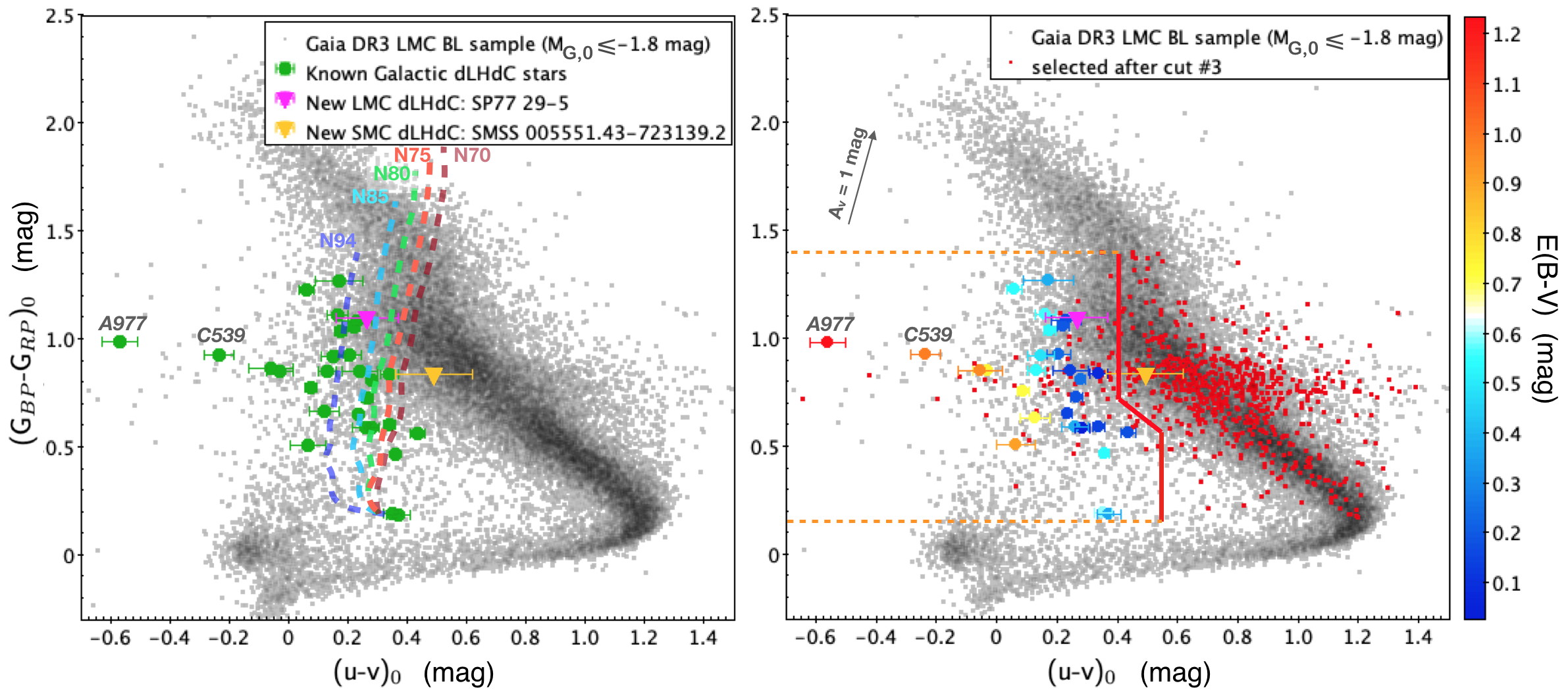}
    \caption{(G$_{BP}$-G$_{RP}$)$_0$ vs (u-v)$_0$ diagram with the same legend as fig.~\ref{fig:HRFig}. Left: The coloured dashed lines indicate the expected position of dLHdC stars based on synthetic spectra made from H-deficient MARCS model for different N abundances in the 4000 - 7500 K effective temperature range. Right: distribution in red of the selected BL stars after cuts \#2 + \#3. The solid red lines illustrate the selection applied with cut \#4 and the orange dashed lines represent the colour selection previously applied to (G$_{BP}$-G$_{RP}$)$_0$ with cut \#2. The known Galactic dLHdC stars are colour coded with their E(B-V) reddening values. Some, such as A977 and C539 indicated in the diagrams, that are the most impacted by interstellar extinction, show bluer values than expected from the synthetic spectra. This is most likely due to the well-documented SkyMapper u-band red leak (see text).}
    \label{fig:BpRpvsuv}
\end{figure*}

\begin{table*}[htbp]
    \centering
    \begin{tabular}{lrr}
\hline
         & LMC    & SMC  \\
\hline
Total number of Gaia BL stars  & 261929  &  86212  \\
With IR counterpart:  &  224260 / 14986 & 79143 / 2309 \\
With E(B-V)$\leq$0.2 mag: &  222631 / 14898 & 79138 / 2309\\
Selection \#1: (J-Ks)$_0$ vs (G$_{BP}$-G$_{RP}$)$_0$ & 3401 / 2093  & 1102 / 250  \\
Selection \#2: HR diagram  M$_{G,0}$ vs (G$_{BP}$-G$_{RP}$)$_0$    & 915 / 73  & 324 / 26  \\
Selection \#3: Valid u \& v SkyMapper DR4 photometry        & 631 / 57 &  139 / 7  \\
Selection \#4: (G$_{BP}$-G$_{RP}$)$_0$ vs (u-v)$_0$ &  73 / 7  &  30 / 2  \\
\hline
After SIMBAD (known variable stars removed)       & 12 / 5   & 5 / 2  \\
\hline
    \end{tabular}
    \caption{Number of selected Gaia DR3 BL stars with VMC DR6 / 2MASS-6X IR counterpart after each selection cut.}
    \label{tab:selectioncuts}
\end{table*}

Third, we selected those presenting valid PSF photometric measurements in both SkyMapper DR4 u and v bands. About 70\% and 40\% of the remaining BL stars in the LMC and SMC, respectively, presented such valid information. Among them, we applied a fourth selection cut, keeping the ones located within an area defined in the (G$_{BP}$-G$_{RP}$)$_0$ versus (u-v)$_0$ diagram (see Fig.~\ref{fig:BpRpvsuv}). Because HdC stars lack hydrogen in their atmospheres, their spectra do not display a Balmer jump, making them stand out remarkably. Furthermore, synthetic spectra show that the higher the nitrogen abundance, the bluer the HdC stars, due to a prominent CN bandhead located within the v band. We applied the following selection cuts and picked 112 Magellanic stars (80 and 32, respectively, in the LMC and SMC):

\begin{equation}
if\hspace{0.1cm}(G_{BP}-G_{RP})_0\geq0.72 : \hspace{0.2cm}(u-v)_0 \leq 0.4
 \label{eq_uv_BPRP_a}
\end{equation}

\begin{equation}
\begin{array}{l}
if\hspace{0.1cm}0.58\leq(G_{BP}-G_{RP})_0<0.72 : \\
              \hspace{3.7cm}(u-v)_0 \leq -(G_{BP}-G_{RP})_0 + 1.12
\end{array}
\label{eq_uv_BPRP_b}
\end{equation}

\begin{equation}
if\hspace{0.1cm}(G_{BP}-G_{RP})_0<0.58 : \hspace{0.2cm}(u-v)_0 \leq 0.54
 \label{eq_uv_BPRP_c}
\end{equation}

Among the known Galactic dLHdC reference stars, four of them -- A977, C539, B563, and B566 -- show dereddened $(u-v)_0$ colours with negative values, between 0.2 and 0.7 mag bluer than predicted by our synthetic grid. Overall, we are observing that the size of the discrepancy increases monotonically with reddening, particularly for dLHdC stars with $\mathrm{E(B-V)}>0.6$ mag (see Fig.~\ref{fig:BpRpvsuv}, right). We investigated and excluded several explanations. Contamination by blended sources is excluded by Gaia DR3: cone searches reveal no neighbour capable of supplying the required $u$-band flux, and the stars' own image-parameter diagnostics ($\mathtt{ipd\_frac\_multi\_peak}=0$, $\mathrm{RUWE}<1.5$) exclude unresolved companions. Instead, the anomaly is naturally explained by the documented red leak of the SkyMapper $u$ filter at 700-750 nm \citep{2011Bessell, SkyMapperDR4-Onken2024} at the one-percent level. For a cool carbon star behind $E(B-V) \approx 1$, the photospheric $u$-band flux is suppressed by $A_u\sim5$--$6$ mag while the flux feeding the leak is attenuated by only $\sim$2 mag, so the leak contribution -- negligible for the bulk of our sample -- can approach the photospheric signal itself, brightening $u$ by several tenths of a magnitude. This mechanism accounts simultaneously for the amplitude of the anomaly, its restriction to the $u$ band, its repeatability across the individual SkyMapper epochs, and its scaling with $E(B-V)$. 

After the application of the four selection criteria, a cross-match of the 112 selected stars with the SIMBAD database revealed that 83 of them are Cepheids and two others are RR Lyrae variables. Their photometric variability in luminosity and colour allowed them to pass the previous selection criteria due to the non-simultaneity of the observations between the surveys. In most cases, their classifications resulted from identifications made by the OGLE monitoring survey  \citep{Soszynski2008,2015AcA....65....1U}. We checked their publicly available light curves and also rejected one clear eclipsing binary object. Furthermore, two known Magellanic RCB stars, namely EROS2-LMC-RCB-5 \citep{Tisserand2009} and ASAS J050232-7218.9 \citep{2014JAVSO..42...13O}, remained present at that stage in our selection and were therefore removed  (we return to the significance of their recovery in Sect.~\ref{sec:detecteff}). Finally, we ended up with 17 and 7 stars of interest, respectively, in the LMC and SMC. They do not have specific reported classification in the SIMBAD database. 

The spectroscopic follow-up of each selected star of interest allowed the discovery of the first dLHdC star located in the LMC, named SP77 29-5 (see Sect.~\ref{sec:SP77_29-5} and the discussion therein). All other targets show characteristics of hydrogen-rich stars and many of them are foreground stars.

\subsection{Method 2: Machine learning random-forest algorithm \label{sec:RF}}

\begin{figure*}[htbp]
    \centering
        \includegraphics[width=1.0\linewidth]{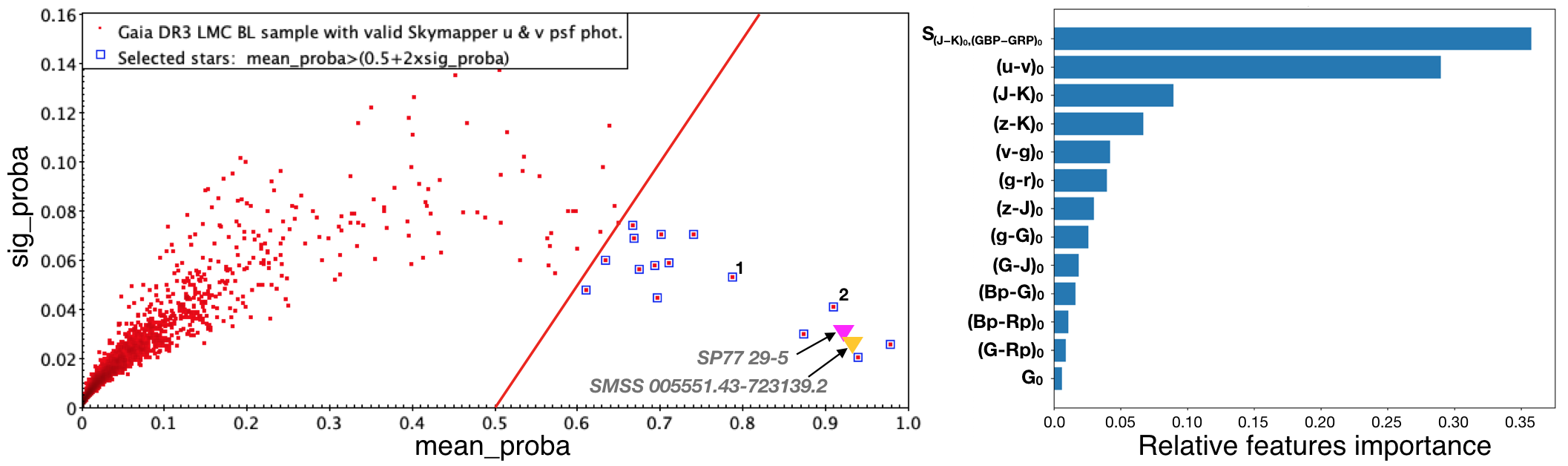}
        \caption{Left: diagram of the probability standard deviation versus the mean probability for all Gaia DR3 LMC BL stars with valid SkyMapper u and v photometry, based on random-forest algorithms. The selection cut is illustrated with a red line and the selected stars for spectroscopic follow-up are indicated with blue squares. The two newly discovered dLHdC stars are represented with coloured down triangles. The stars labelled with the numbers 1 and 2 are respectively the known Magellanic RCB stars ASAS J050232-7218.9 and EROS2-LMC-RCB-5. Right: horizontal bar chart representing the relative importance between all features used by the random-forest algorithm to discriminate between background hydrogen-rich stars and dLHdC stars.}        
    \label{fig:RF_Proba_Features}
\end{figure*}

\subsubsection{Training strategy and validation \label{sec:RFtraining}}

We turn to machine-learning classification techniques, and in particular to the random-forest (RF) algorithm from the scikit-learn Python package \citep{Pedregosa2011_Scikit-learn}, to search for Magellanic dLHdC stars. A random forest is an ensemble of decision trees, each trained on a bootstrap resampling of the data and allowed to split on a random subset of the features at every node; the class probability assigned to a source is the average over all trees \citep{Breiman2001}. Because dLHdC stars are extremely rare, classical selection cuts could become inefficient in colour--colour regions that are heavily contaminated by background populations, especially near the limits of photometric precision. RF classification offers a way to exploit multidimensional photometric information simultaneously and thereby improve our detection efficiency in these challenging regions. Indeed, RF captures non-linear boundaries in a multidimensional colour space without requiring them to be specified in advance, unlike our classical cuts; it is robust, in terms of classification performance, to the strong correlations that exist between colour indices constructed from overlapping passbands; it provides a natural ranking of the discriminating power of each photometric feature; and finally, its output probabilities can be turned into a quantitative selection threshold.

For each experiment, the RF classifier was trained using two labelled samples: (i) the target class (dLHdC stars) and (ii) typical Magellanic background stars. Each training sample contained 10000 objects, reduced to 5000 when the parent Magellanic sample was small ($\leq$20 000 stars), to preserve representativeness. For the target class, we randomly drew stars from the small set of known Galactic dLHdC stars, allowing objects to be sampled multiple times. For the background class, stars were randomly selected from the Gaia Magellanic BL subsample, or even a selection of them in colour and luminosity. The RF algorithm relies on a collection of features describing each source, mostly colour indices constructed from Gaia, SkyMapper and VMC photometry. These features were selected after examining their relative importance and impact on classification performance. Feature importances were computed using the mean decrease in impurity (the default in scikit-learn), i.e. the total reduction of the entropy criterion contributed by each feature, averaged over all trees of the ensemble and normalised to unity. The complete list of features used in each analysis is provided in the following subsections.

To increase the diversity of the training data, we applied a data-augmentation procedure that perturbs the photometry according to the reported measurement uncertainties. For each photometric band, a new magnitude was drawn from a Gaussian distribution centred on the catalogue value with a standard deviation equal to the quoted photometric error. We also introduced variations in the extinction correction by adopting Gaussian uncertainties of 0.10 and 0.03\,mag in $E(B-V)$ for the known Galactic targets and the Magellanic background stars, respectively. When intrinsic luminosities were used as features, we further applied data augmentation to the distances of the known Galactic dLHdC stars. The minimum and maximum geometric distances from \citet{Bailer-Jones2021} were taken as a $\pm 1\sigma$ interval around the median value, and new distances were drawn accordingly. This procedure mitigates overfitting to the limited set of known objects and allows the RF classifier to explore a more realistic range of photometric and physical properties.

We adopted a 5-fold cross-validation strategy in which the full training set is randomly shuffled and partitioned into five subsets. One fold is used once for validation while the remaining four are used for training, and this process is repeated independently for every RF experiment. Performance metrics (accuracy, precision, recall, and F1-score), averaged over the five folds, consistently exceeded 99\% when applying a probability threshold of 0.6 for membership in the target class. We used the \texttt{entropy} criterion to evaluate the quality of each split and imposed no maximum tree depth or minimum number of samples required for node splitting. No explicit class weighting was applied. The confusion matrices showed negligible contamination, confirming that the RF algorithm effectively separates the target population from the dominant Magellanic background. 

For each analysis, to mitigate stochastic variability and to obtain uncertainty estimates on output probabilities, we trained an ensemble of 100 independent RF classifiers, each composed of 500 trees. Combining predictions from models differing only in their random initialisation has been shown to yield more robust and stable classifications than any single model in astronomical applications \citep[e.g.][]{Moller2022}. For each Magellanic star, every classifier produced a probability of belonging to the target class. For each star, we then computed the mean and the standard deviation of the resulting probability distribution---hereafter \texttt{mean\_proba} and \texttt{sig\_proba}---which provide both a central likelihood estimate and a measure of the robustness of the classification across the ensemble.

Finally, the sample of Magellanic background stars used for training may itself contain genuine dLHdC stars, which would degrade the classifier by teaching it to assign true targets to the background class. To limit this effect, the analysis was performed in two passes: stars with a \texttt{mean\_proba} value higher than 0.5 in the first pass were removed from the training set used in the second pass. Only a few stars were put aside, i.e., about 0.2\% of the Magellanic background stars. The main effect on the second pass classification is a reduction by a factor of 2--3 in the classifier-to-classifier dispersion \texttt{sig\_proba} for these objects of higher probability, i.e. a more stable classification of precisely the stars of interest.


\subsubsection{Application to the Magellanic sample \label{sec:RFapplication}}

We focused our RF analysis on Magellanic stars belonging to the BL subsample whose colours ($0.1 < (G_{BP} - G_{RP})_0 < 1.5$ mag, slightly wider range than for the classical selection) and intrinsic brightnesses ($-5.5 < M_{G,0} < -1.8$ mag) are consistent with those of known Galactic dLHdC stars. After removing the stars that did not have an IR counterpart (from VMC DR6 or 2MASS-6X) or were affected by high interstellar reddening ($E(B-V)>$0.2 mag), this selection yields a total of 46\,188 and 23\,719 stars in the LMC and SMC, respectively. By construction, this approach excludes the more numerous fainter stars, which would otherwise unnecessarily bias the classification algorithm. The adopted faint-magnitude limit also approximately matches the sensitivity limits of the SkyMapper $u_{\mathrm{PSF}}$ and $v_{\mathrm{PSF}}$ photometry. Consequently, among the pre-selected Magellanic stars, about half (21\,734 stars) in the LMC and one third (8\,308 stars) in the SMC have reliable measurements in these blue passbands. We therefore developed two distinct RF analyses, using feature sets that either include or exclude the SkyMapper $u_{\mathrm{PSF}}$ and $v_{\mathrm{PSF}}$ magnitudes. They are named respectively RF$_{UV}$ and RF$_{NoUV}$.

Our first RF analysis, RF$_{NoUV}$, which does not use the $u_{\mathrm{PSF}}$ and $v_{\mathrm{PSF}}$ data, was applied to all pre-selected Magellanic stars using a set of 11 features, listed here in increasing order of importance: $(G - G_{RP})_0$, $M_{G,0}$, $(G_{BP} - G_{RP})_0$, $(G_{BP} - G)_0$, $(g_{\mathrm{PSF}} - G)_0$, $(G - J)_0$, $(g_{\mathrm{PSF}} - r_{\mathrm{PSF}})_0$, $(z_{\mathrm{PSF}} - J)_0$, $(z_{\mathrm{PSF}} - K_s)_0$, $(J - K_s)_0$, and a custom parameter, $S_{(J-K_s)_0,(G_{BP}-G_{RP})_0}$. This last parameter represents the minimum distance, in the $(J-K_s)_0$ versus $(G_{BP}-G_{RP})_0$ diagram, to the empirical separation line defined in Eq.~\ref{eq_JK_BPRP_a} between known dLHdC stars and background populations. In a second analysis, RF$_{UV}$, we added the $(v_{\mathrm{PSF}} - g_{\mathrm{PSF}})_0$ and $(u_{\mathrm{PSF}} - v_{\mathrm{PSF}})_0$ colours to the feature set. As expected, the $(u_{\mathrm{PSF}} - v_{\mathrm{PSF}})_0$ colour emerged as the second most important feature in the classification (see Fig.~\ref{fig:RF_Proba_Features}, right). Note that the features made from the Gaia DR3 wide passbands are of the lowest importance to discriminate between background (hydrogen-rich) stars and dLHdC stars.

Candidates were retained if they satisfied the following criteria: (i) an absolute magnitude
$M_{G,0} < -2$ mag, and (ii) a mean classification probability exceeding 0.5 by more than two standard deviations, i.e.\ \texttt{mean\_proba} $- 2 \times$ \texttt{sig\_proba} $\geq 0.5$ (see Fig.~\ref{fig:RF_Proba_Features}, left), with RF$_{UV}$ and by more than three standard deviations with RF$_{NoUV}$. In that last analysis, among the selected objects we rejected the ones presenting clear signs of large photometric variability using the (g-G)$_0$ parameter and/or, if available, the (u-v)$_0$ colour index (red objects with values higher than 0.8 mag were rejected). This conservative selection ensures the identification of objects with both high likelihood and low classifier-to-classifier variance.

In total, we selected 45 and 19 candidates, respectively, in the LMC and SMC. Cross-matching with the SIMBAD database revealed that 31 of these objects are identified as Cepheids, two others as long-period variables and one as an RR Lyrae star; they were consequently discarded. Finally, 19 (11) candidates were retained for spectroscopic follow-up in the LMC (SMC), of which only 10 (2) had already been identified in the classical analysis. Interestingly, among them, we again recover the two known RCB stars, ASAS\,J050232-7218.9 and EROS2-LMC-RCB-5, and the newly identified Magellanic dLHdC star SP77 29-5 (see Sect.~\ref{sec:SP77_29-5}). SP77 29-5 yielded mean probabilities of $0.92 \pm 0.03$ in RF$_{UV}$ and $0.999 \pm 0.002$ in RF$_{NoUV}$.

All other selected targets that were followed up spectroscopically were found to possess hydrogen-rich atmospheres, except for one star located in the SMC. It is named SMSS 005551.43-723139.2 (hereafter SMSS0055-72) and is the first dLHdC star discovered in that galaxy. It was reported with mean probabilities of $0.93 \pm 0.03$ in RF$_{UV}$ and $0.998\pm0.002$ in RF$_{NoUV}$. SMSS0055-72 was not selected by the classical analysis because it presents a high (u-v)$_0$ colour value (see cut \#4 in Fig.~\ref{fig:BpRpvsuv}, right). It is at the limit of detectability in the SkyMapper u band, with $u_{\mathrm{PSF}}=17.1\pm0.12$ mag. The characteristics of SMSS0055-72 are discussed in detail in Sect.~\ref{sec:SMSS0055}.

\subsection{Detection efficiency \label{sec:detecteff}}

Our primary selection cuts were defined to encompass all known Galactic dLHdC stars used as references in our analysis. Only two of them, C542 and A814, did not pass the full chain of selection criteria, specifically cut~\#3, as they lacked valid photometric measurements in the SkyMapper u and v bands. However, their failure results from their specific Galactic circumstances --- a combination of large distances  and high interstellar reddening suppressing their blue-band fluxes ---  conditions that are not representative of our Magellanic search sample, which was restricted to low reddening values, E(B-V)~$\leq$~0.2~mag. Furthermore, Magellanic stars lacking valid u and v photometry and thus excluded by classical cut~\#3 (i.e.,  $\sim$30\% and $\sim$60\%\footnote{These fractions refer to the full BL population; the intrinsically more luminous and bluer dLHdC stars are better represented in the SkyMapper blue passbands (see Fig.~\ref{fig:HRFig}, left).} of the remaining stars in the LMC and SMC, respectively, see Table~\ref{tab:selectioncuts})  remain recoverable by the RF analysis based on the feature set that excludes these bands (Sect.~\ref{sec:RFapplication}). Their exclusion at cut~\#3 therefore does not bias our overall detection efficiency estimate.

We estimated the overall detection efficiency of our combined selection by processing the data-augmented sample of known Galactic dLHdC stars (Sect.~\ref{sec:RFtraining}) --- realisations that probe the photometric and physical parameter space around the known objects rather than the objects themselves --- placed at the LMC and SMC distance moduli, through the full chain of colour and luminosity selection criteria and RF classifications. Overall, 96\% of the simulated Magellanic dLHdC stars are recovered by at least one of the two methods, while the classical selection alone already reaches $\sim$86\% in the LMC and $\sim$81\% in the SMC, once the loss of objects lacking valid SkyMapper u and v photometry (cut~\#3) is taken into account.

The discovery of SMSS0055-72 provides an empirical validation of the redundancy between our two analyses: lying at the detection limit of the SkyMapper u band (u$_{\mathrm{PSF}}$ = 17.1~$\pm$~0.12~mag), it failed selection cut~\#4 in the classical analysis, but was recovered with high confidence by the RF classification (Sect.~\ref{sec:RFapplication}). The two methods are thus complementary precisely in the regime where the blue-band photometry becomes unreliable.

The recovery of the two known RCB stars, ASAS J050232-7218.9 and EROS2-LMC-RCB-5, by both the classical and RF analyses provides a further empirical validation of our method. These are the only two Magellanic RCB stars that did not undergo a photometric decline during the Gaia DR3 observation window, as confirmed by their OGLE-IV light curves \citep[see their RCOM website]{Udalski2008}\footnote{\url{https://ogle.astrouw.edu.pl/ogle4/rcom/rcom.html}}, and that were, at the same epochs, free of any warm circumstellar dust excess in their NEOWISE light curves\footnote{The NEOWISE light curves show a flat W1$-$W2 colour consistent with pure photospheric emission for EROS2-LMC-RCB-5, and, for ASAS J050232-7218.9, a colour ($W1-W2\sim1.6$ mag) indicating a cold shell whose emission is negligible at J and K$_s$.}. Both their optical and near-IR photometry therefore correspond to a quiescent, uncontaminated HdC-type spectral energy distribution, in contrast to the other known Magellanic RCB stars, whose mean photometry is displaced from that locus by
decline epochs and/or dust emission. Our selection thus recovered every star in the sample presenting such an SED -- the regime in which dLHdC stars, which never fade, are always found.

Finally, we note that the last selection cut~\#4 kept all the known Galactic dLHdC stars, but a comparison with the photometry resulting from the synthetic models shows that cool dLHdC stars (T$_{eff}\leq5750$ K) of low nitrogen abundance ($[N]\leq7.5$) would be missed (see Fig.~\ref{fig:BpRpvsuv}). Such dLHdC stars are not represented in the current sample of known Galactic dLHdC stars. Our quoted efficiency therefore applies to dLHdC stars with properties similar to the known Galactic population; a hypothetical population of cool, nitrogen-poor dLHdC stars would be underrepresented, a caveat consistent with the discussion of selection effects in Sect.~\ref{sec:discus_1}.

\begin{figure*}[ht]
    \centering
    \includegraphics[width=1.0\linewidth]{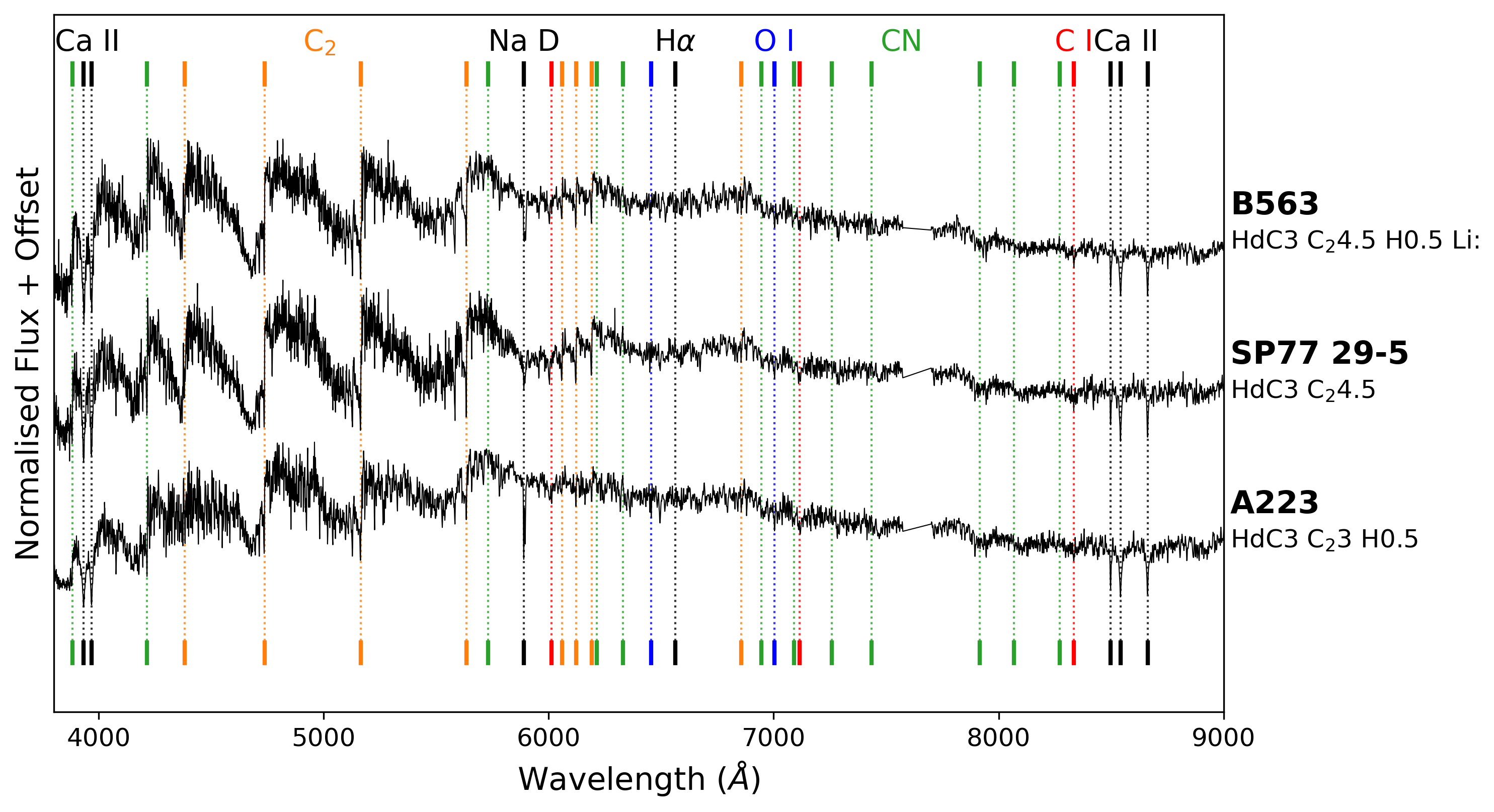}
    \caption{Optical spectra taken with the 2.3m/WiFeS spectrograph of B563, SP77 29-5 and A223. They are all dLHdC stars of the same effective temperature, classified as HdC3 \citep{Crawford2023}. The main chemical signatures are indicated with coloured vertical dotted lines.}
    \label{fig:SP77_29-5_Spectrum}
\end{figure*}

\section{Newly found Magellanic dLHdC stars \label{sec:New-dLHdC}}

Our two analyses led us to discover the first Magellanic dLHdC stars, namely SP77 29-5 and SMSS 005551.43-723139.2, whose main properties are listed in Table~\ref{tab:NewdLHdC}. Their characteristics are discussed hereafter. 

\begin{table*}[]
    \centering
    \begin{tabular}{lccccl}
\hline
   Name   & Coordinates    & Gaia DR3  &  G & (G$_{BP}$-G$_{RP}$)$_0$& HdC spectral class  \\
      & (J2016)    & source id  &   (mag)  & (mag)  & \\
\hline
 SP77 29-5   & 04:54:21.66 -66:34:48.33 &  4662158382480976640 & 14.87 & 1.10 &  HdC3 C$_2$ 4.5 \\
 SMSS 005551.43-723139.2 & 00:55:51.44 -72:31:39.16  & 4688990275885108480 & 15.32 & 0.83 & HdC2 C$_2$ 3.5 H1 CN- Sr \\
\hline
    \end{tabular}
    \caption{Summary table of the two new Magellanic dLHdC stars.}
    \label{tab:NewdLHdC}
\end{table*}

\subsection{SP77 29-5, the first dLHdC star in the LMC \label{sec:SP77_29-5}}

SP77 29-5 is located on the edge of the LMC, near the north-western spiral structure. It was reported by the Gaia DR3 survey with a radial velocity of $278.74\pm2.62$ km s$^{-1}$, and with proper motion values within 1$\sigma$ of the distribution of LMC BL stars located within 10 arcmin radius. Its kinematics confirm its LMC membership. SP77 29-5 was first mentioned by \citet{Sanduleak1977} in their catalogue of suspected carbon-star supergiants in the LMC. Based on their low-dispersion spectrum, \citet{Hartwick1988} confirmed SP77 29-5 as a carbon star and rejected it from their initial list of CH stars. Later, \citet[number 897]{Kontizas2001} conducted a systematic survey of the LMC using objective-prism plates and validated the classification of SP77 29-5 as a carbon-rich star. With 2MASS magnitudes of K$\sim$13.3 mag and J-K$\sim$0.4 mag, SP77 29-5 already stood out from the usual locus of Magellanic carbon rich stars in the presented IR colour-magnitude diagram (see their Fig. 5).

\begin{figure}[h]
    \centering
    \includegraphics[width=1.0\linewidth]{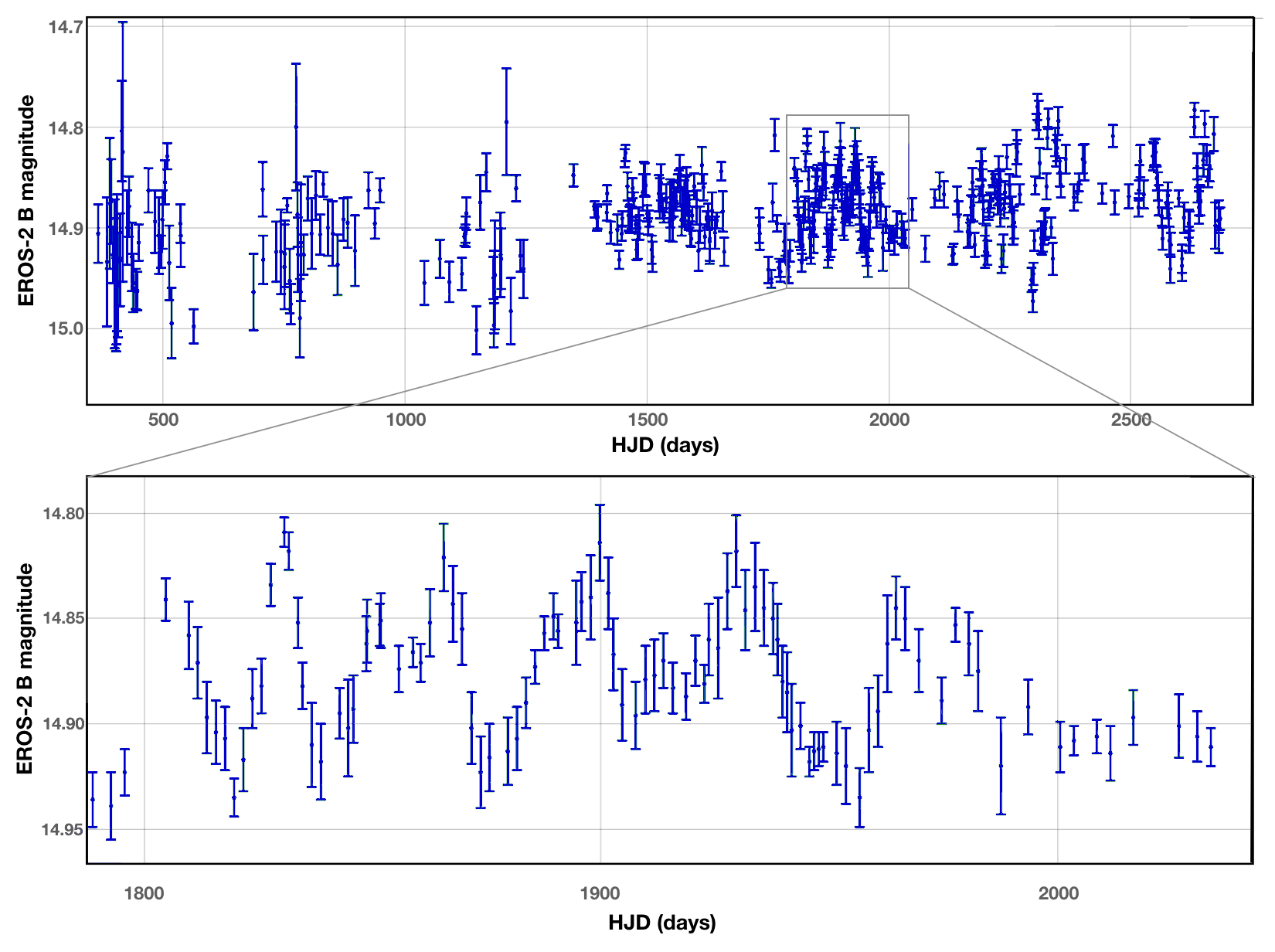}
    \caption{EROS-2 B band light curve of SP77 29-5. Top: entire light curve between 1996 and 2003. Bottom: Zoom on the $\sim$0.12 mag amplitude variations observed between HJD 1800 and 2000 days.}
    \label{fig:EROS2_LC}
\end{figure}

The OGLE monitoring survey labelled it as a variable star, OGLE-LMC-LPV-06252, with a primary period of 17.24 days and a weak peak-to-peak amplitude of $\sim$0.02 mag in the I band \citep{Soszynski2009}. It was then classified as an OSARG (acronym for OGLE Small Amplitude Red Giant). However, our visual inspection of its EROS-2 light curve \citep{EROS2-2026} reveals semi-regular variations of $\sim$0.12 mag peak-to-peak amplitude in both the B and R bands, with a periodicity comparable to the 17.24 day period reported by OGLE (Fig.~\ref{fig:EROS2_LC}). This is typical of the photometric behaviour of dLHdC stars.

SP77 29-5's HR diagram position (M$_V\sim-3.5$ mag and (V-I)$_0\sim$1.0 mag from OGLE survey) shows that it is about 1.0 mag fainter than Magellanic RCB stars of the same colour, but still at the bright end of the known Galactic dLHdC distribution \citep[see their Fig.10, bottom]{Tisserand2022}. Its SkyMapper (u-v)$_0\sim$0.26 mag colour together with the following two colour indices, (G$_{BP}$-G$_RP$)$_0\sim$1.1 mag and $(J_\mathrm{aper3}-K_{\mathrm{s,aper3}})_0\sim$0.36 mag, indicate that SP77 29-5 is indeed missing the Balmer jump feature. SP77 29-5 is located within the locus of known dLHdC stars in colour-colour diagrams (see Figs.~\ref{fig:JKvsBpRp} and ~\ref{fig:BpRpvsuv} for direct comparison).

\begin{figure*}[!htp]
    \centering
    \includegraphics[width=1.0\linewidth]{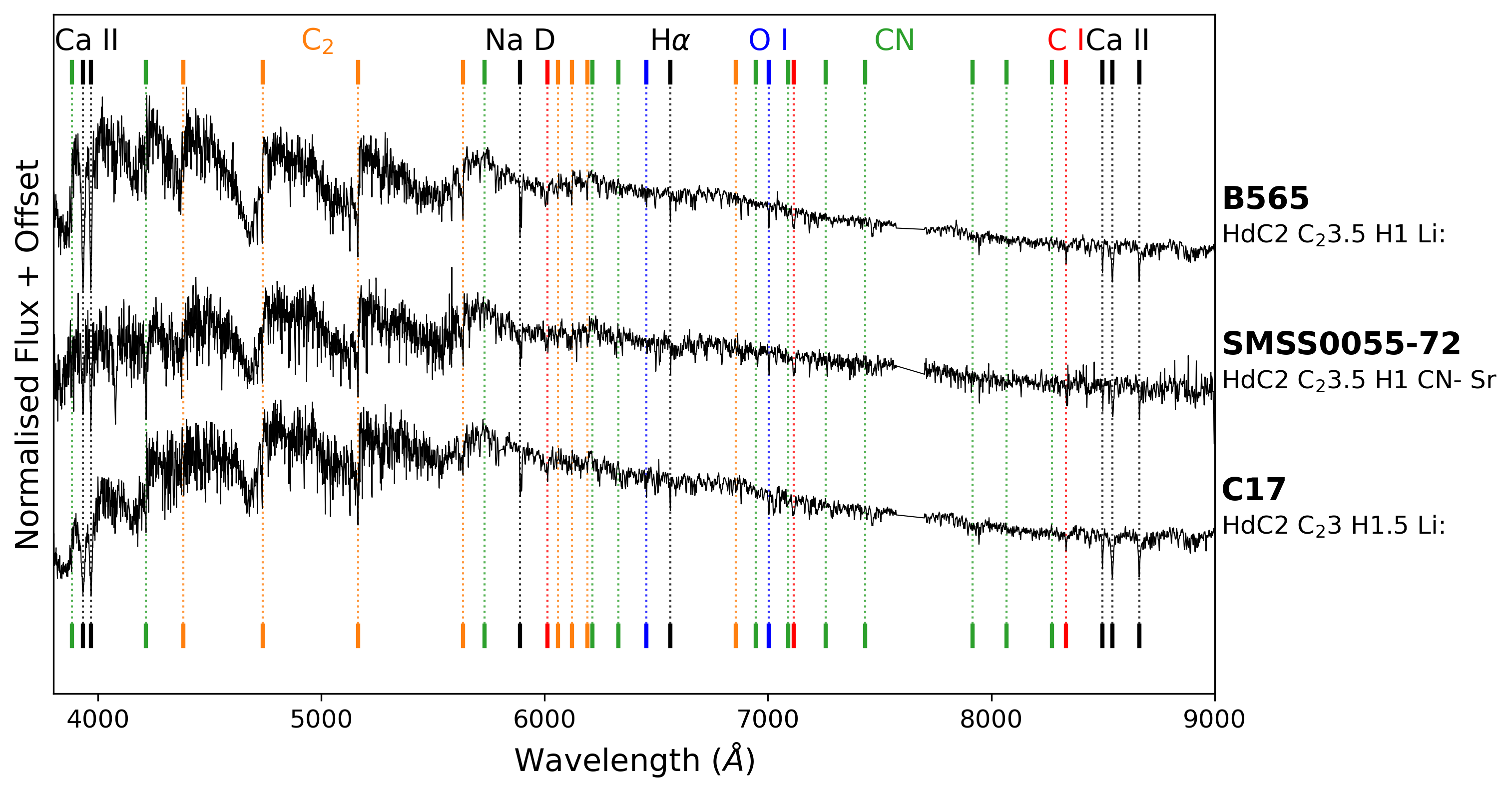}
    \caption{Optical spectrum of SMSS 005551.43-723139.2 taken with the 2.3m/WiFeS spectrograph over the entire wavelength range with a resolution R$=$3000, with two other Galactic dLHdC stars, B565 and C17, of the same effective temperature. Coloured vertical lines indicate the main atomic absorption lines (Ca II , Na D, O I, C I and H$\alpha$) and band-heads of features due to C$_2$ and CN molecules.}
    \label{fig:SMSS0055-72_spectrum}
\end{figure*}

We obtained a medium-resolution spectrum with the 2.3m/WiFeS spectrograph on 2022 February 12 (R$\sim$3000, total exposure time of 900 s, Fig.~\ref{fig:SP77_29-5_Spectrum}) that confirmed the hydrogen deficiency of  SP77 29-5's atmosphere with no sign of Balmer lines nor of the CH molecular band at $\sim$4300\AA. Using the classification system developed by \citet{Crawford2023}, which classifies HdC stars between the warmest (HdC0, T$_{eff}\sim$8750 K) and the coolest (HdC7, T$_{eff}\sim$4000 K), we labelled SP77 29-5 with a spectral type of HdC3 C$_2$ 4.5, with the last value presenting the strength of the C$_2$ carbon band. The spectral type indicates a medium temperature range ($\sim$5500 - 6000 K). SP77 29-5's spectrum is very similar to that of the Galactic dLHdC star B563 that has an identical spectral class. It also has C$_2$ features slightly stronger than those of A223, which shares the same spectral type.  

\begin{figure}[h]
    \centering
    \includegraphics[width=1.0\linewidth]{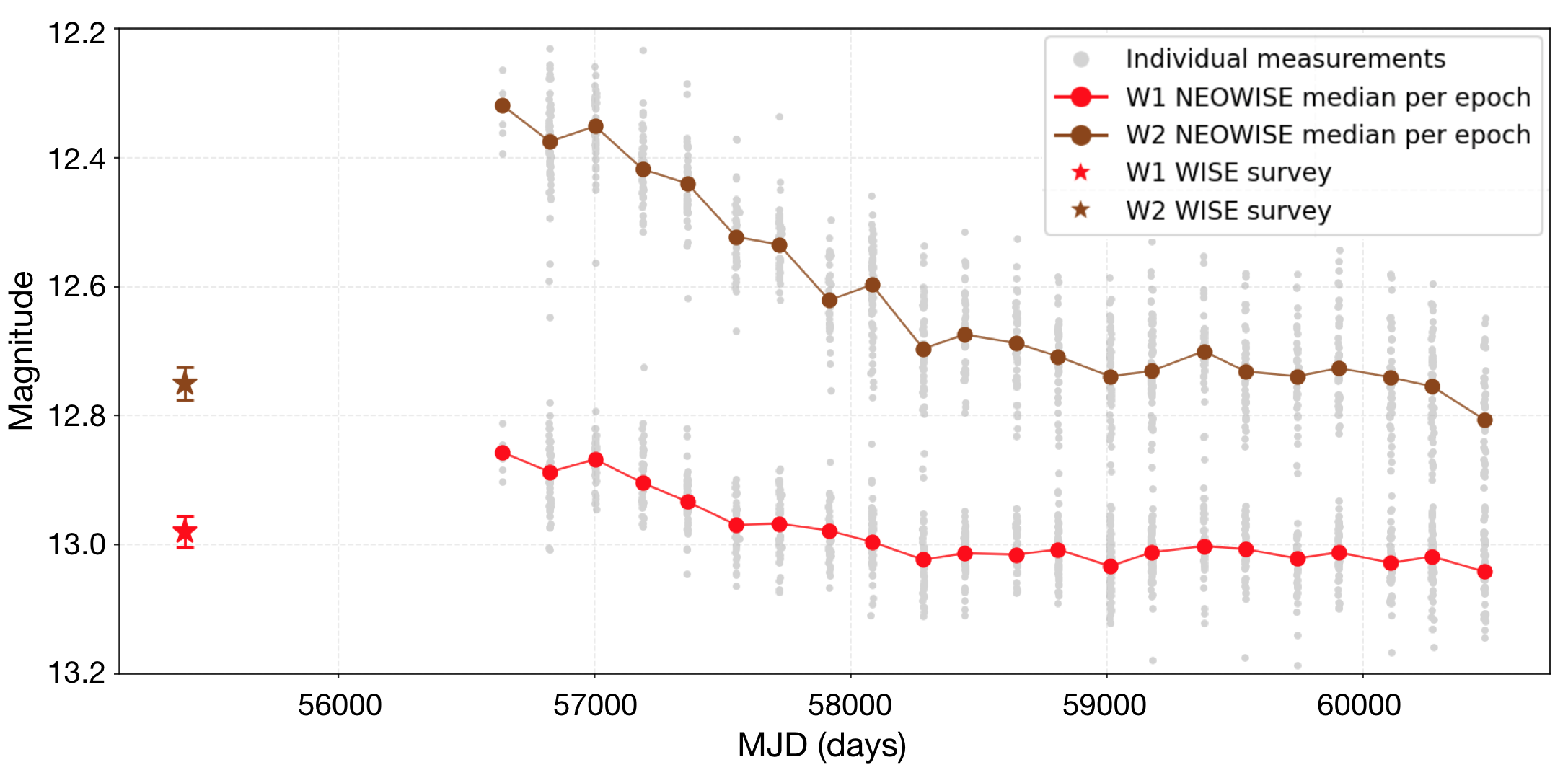}
    \caption{SP77 29-5 mid-IR NEOWISE W1 and W2 light curves. Individual NEOWISE measurements are in grey while the median in each narrow time window is indicated in red for W1 and in maroon for W2. The ALLWISE measurements with their associated error bars obtained during the 2010 survey are indicated with stars.}
    \label{fig:NEOWISE_lc}
\end{figure}

The SP77 29-5 NEOWISE light curve (2013-2024, see Fig.~\ref{fig:NEOWISE_lc}) presents variations of $\sim$0.1 and $\sim$0.4 mag in the W1 and W2 bands, respectively, indicating a low dust production rate similar to that observed for four Galactic dLHdC stars, i.e. F75, F152, C526 and A166 \citep{Tisserand2022}.

\subsection{SMSS 005551.43-723139.2, the first SMC dLHdC star \label{sec:SMSS0055}}

SMSS0055-72 has not previously been reported in the literature. It is located only 20 arcmin from the photometric centre of the SMC. Although Gaia DR3 did not provide a radial velocity measurement, we measured a value of $170 \pm 30$ km s$^{-1}$, consistent with membership in the SMC. Furthermore, we checked the proper motions of SMC BL stars located within 10 arcmin around SMSS0055-72. SMSS0055-72 lies within $\sim$1$\sigma$ of the proper-motion ($\mu_{RA}$ versus $\mu_{Dec}$) distribution.

The Gaia DR3 astrometric solution appears robust, with quality indicators of $\mathtt{RUWE}=1.187$, $\mathtt{ipd\_gof\_harmonic\_amplitude}= 0.016$, $\mathtt{RFG}=1056$ (RFG corresponding to the G-band flux-to-error ratio). Although 18\% of the successful IPD\footnote{IPD = Image Parameter Determination} windows contain multiple peaks, this is consistent with the presence of a neighbouring source located $1.2''$ away and approximately 2.9 mag fainter. We therefore find no evidence that this neighbour significantly biases the Gaia DR3 astrometric solution.

SMSS0055-72 was not identified as a variable star in any of the Magellanic Cloud variability studies based on the long-term OGLE monitoring survey, and we detect no obvious variability in the EROS-2 light curve, with an upper limit on a peak-to-peak amplitude of approximately 0.05 mag. However, the Gaia DR3 three-band photometric dispersions catalogue described by \citet{2023A&A...677A.137M} indicates significantly larger scatter than expected for a constant source, with variability indices of 5.8, 10.8, and 8.2 in $G$, $G_{\rm BP}$, and $G_{\rm RP}$, respectively (constant stars typically have values $\leq 3$). Thus, some variability does exist, but with an amplitude below 0.05 mag. 

Likewise, the NEOWISE mid-IR light curves show no obvious variability. The ALLWISE survey photometry gives a $W2-W3$ colour of $1.24 \pm 0.50$ mag \citep{2010AJ....140.1868W}, which would normally suggest the presence of cold circumstellar dust. However, most of the unusually large uncertainty is caused by the W3 measurement. A direct inspection of the W3 image confirmed that no source is actually detected in this band. The quoted W3 magnitude should therefore be regarded as unreliable. Only an upper limit is available in W4. 

The OGLE survey reports $V=14.82$ mag and $V-I=0.76$ mag \citep{1998AcA....48..147U}, in good agreement with the SkyMapper DR4 and Gaia DR3 optical photometry (see HR diagram, Fig.~\ref{fig:HRFig}). SMSS0055-72 is warmer than SP77~29-5 but of similar brightness.

Optical spectra of SMSS0055-72 (see Fig.~\ref{fig:SMSS0055-72_spectrum}) were obtained with WiFeS on the ANU 2.3-m telescope on 2025 June 15 using the B3000 and R3000 gratings ($R\sim3000$, 3400--9560~\AA) with a total exposure time of 1800 s. The spectrum is characteristic of a warm carbon star. We estimated an effective temperature of $\sim$6700 K. The star is classified as an HdC2 C$_2$ 3.5 H1 CN- Sr star \citep{Crawford2023}. The C$_2$ Swan bands are prominent, with the (1,0), (0,0), and (0,1) bandheads at 4737, 5165, and 5635~\AA\ clearly detected, together with particularly weak CN red-system bands. High-excitation C~{\sc i} lines are also present in the red arm, including the 8335~\AA\ line and the 9061--9111~\AA\ multiplet. In contrast, the CH G band near 4300~\AA\ is essentially absent, while the Balmer lines are absent or very weak, demonstrating a hydrogen-deficient photosphere. However, a weak H$\alpha$ absorption feature is detected as in other dLHdC stars \citep{Tisserand2022}.

\begin{figure}[!htbp]
    \centering
    \includegraphics[width=1.0\linewidth]{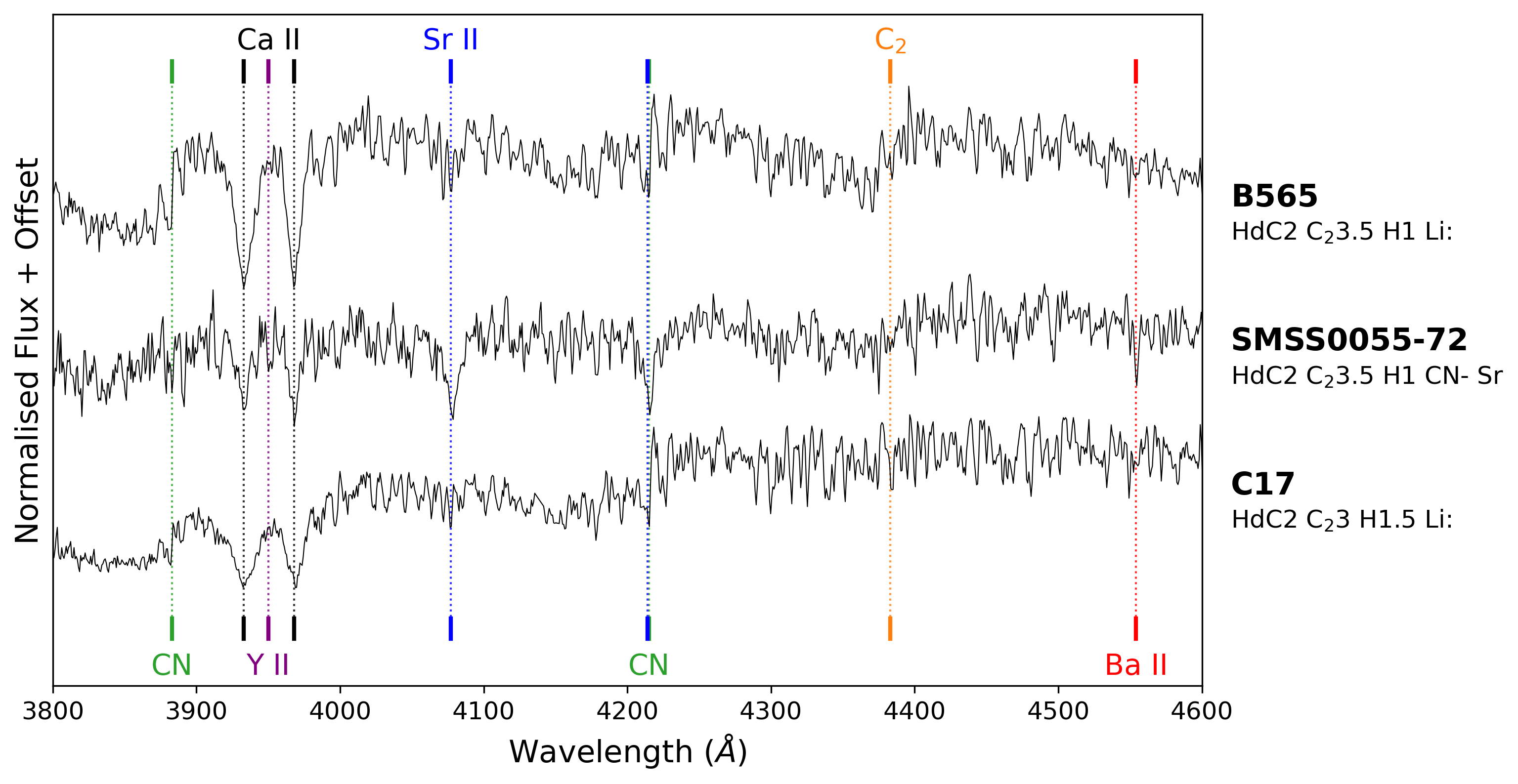}
    \includegraphics[width=1.0\linewidth]{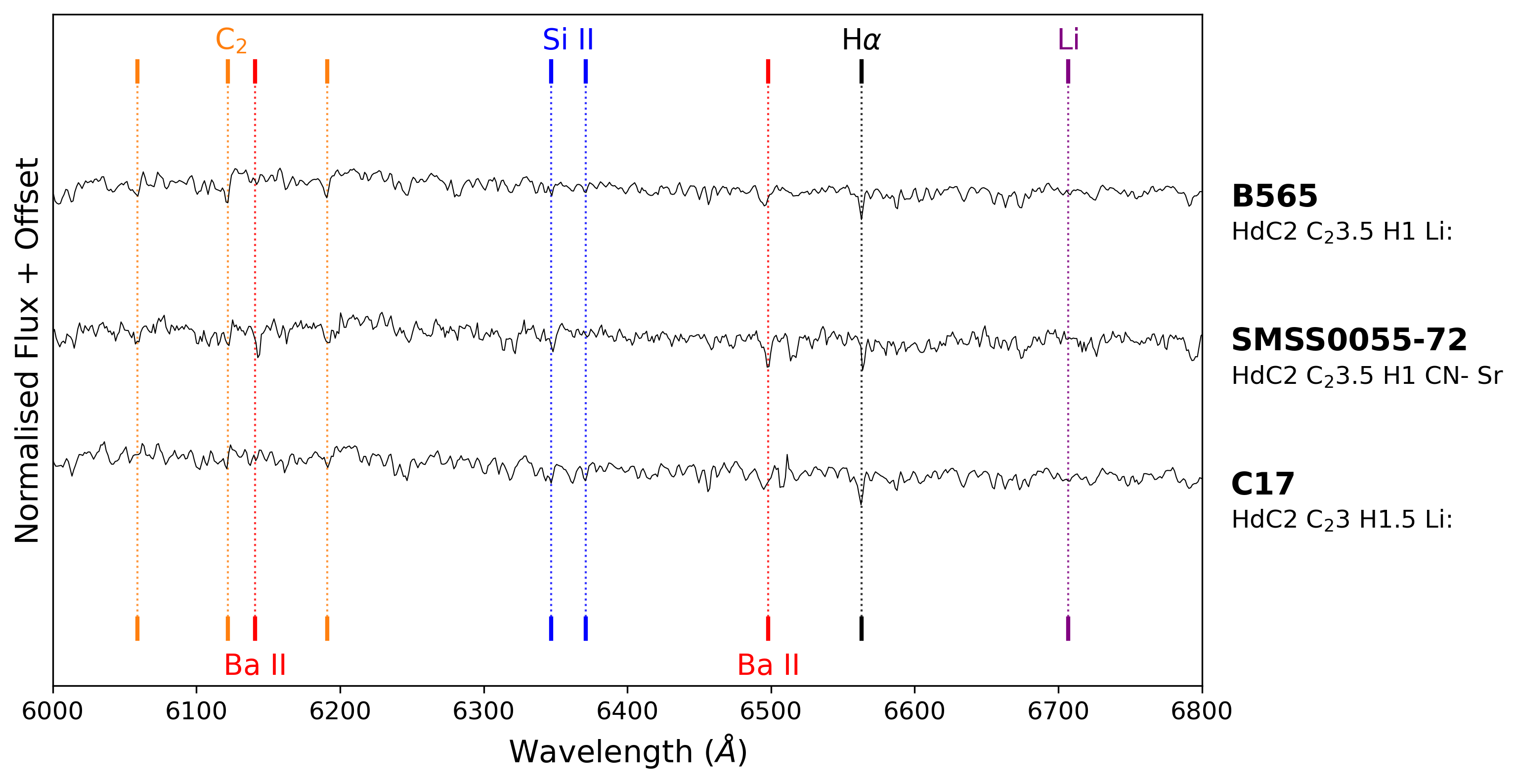}
    \caption{Zoom on the optical spectra of SMSS 005551.43-723139.2 and of two known Galactic dLHdC stars, B565 and C17, in the blue (3800-4600~\AA, top) and in the red (6000-6800~\AA, bottom) wavelength range. Coloured vertical lines indicate the absorption lines observed revealing the strong presence of chemical elements such as Sr and Ba, but also of Y and Si. However, there is no trace of lithium at that resolution. A weak H$\alpha$ line is observable in all three spectra.}
    \label{fig:SMSS0055-72_spectrum_zoom}
\end{figure}

Interestingly, the spectra exhibit the $s$-process signature characteristic of the Sr-rich HdC subclass identified by \citet{Crawford2022}. In Fig.~\ref{fig:SMSS0055-72_spectrum_zoom}, one can observe that the Sr~{\sc ii} $\lambda4077$ line is exceptionally strong, reaching a depth comparable to that of Ca~{\sc ii} H, yielding a Ca~{\sc ii} H/Sr~{\sc ii} $\lambda4077$ pseudo-equivalent-width ratio of approximately 1.0. This is lower than the threshold of 1.5 adopted to define the Sr-rich subclass and far below the typical value of $\sim5.5$ measured in non-Sr-rich HdC stars. The Sr~{\sc ii} $\lambda4215$ line is also detected, blended with the nearby molecular bandhead. Several Y~{\sc ii} lines are present (4374, 4884, 5087, and 5200~\AA), while Ba~{\sc ii} lines (4554, 4934, 6142, and 6497~\AA) are detected but are systematically weaker, in terms of pseudo-equivalent width, than the corresponding Sr and Y features. Also, no lithium absorption line was detected at the observed resolution.

The discovery of an Sr-rich HdC star in the low-metallicity environment of the SMC is particularly interesting. \citet{Crawford2022} noted that most Sr-rich HdC stars are found in the Galactic halo, while the only previously known Magellanic example, EROS2-LMC-RCB-3, also lies in the outer regions of the LMC. This spatial distribution suggests that the Sr-rich subclass is preferentially associated with old, metal-poor stellar populations. The location of SMSS0055-72 is consistent with this picture: unlike in the LMC, where the oldest populations are spatially extended, the old stellar population of the SMC lies near its central regions \citep[][see their Fig.~9]{Rubele2018}. Although this interpretation remains speculative, SMSS0055-72 may thus represent a tracer of one of the oldest stellar populations in the SMC.

\section{Discussion \label{sec:discus}}

We identified only two dLHdC stars in the Magellanic Clouds, whereas 30 RCB stars (24 in the LMC, 6 in the SMC) are currently known in these satellite galaxies. This deficit contrasts with results from dedicated surveys in the Milky Way, where the total number ratio between these two classes of HdC supergiants is estimated to be close to unity \citep{Tisserand2022}. That Galactic value carries a significant uncertainty: the raw ratio measured was closer to one dLHdC star for every two RCB stars, and was corrected upwards principally to account for the shallower effective volume probed for the intrinsically fainter dLHdC stars \citep[][Sect. 4.7]{Tisserand2022}. These corrections can only act to increase the inferred dLHdC population, so that the Galactic ratio is best regarded as a lower limit of order unity -- an order of magnitude above the Magellanic value reported here.

Several factors may contribute to this discrepancy. In the following, we explore possible explanations related to metallicity that may affect the observed dLHdC population through three distinct mechanisms: (i) direct modifications of the atmospheric opacity and effective temperature distribution, (ii) indirect effects on the WD merger mass distribution through binary evolution, and (iii) differences in the durations of the dLHdC and RCB phases themselves. We also examine, in Sect.~\ref{sec:substructures}, whether the relative incidence of the two subclasses already varies among the Galactic stellar populations.

\subsection{Observational selection effects \label{sec:discus_1}}

The effective temperature of the supergiant phase resulting from a double-degenerate merger in the mass regime expected for HdC stars depends on the respective core and envelope masses as shown by \citet{Crawford2024}. 

In their atmospheres, the opacity is strongly influenced by carbon molecules, such that the atmospheric structure depends sensitively on the carbon-to-oxygen ratio. Since HdC star atmospheres typically exhibit C/O ratios far above unity, often of the order of $\sim$30, most oxygen is expected to be locked into CO molecules, while the excess carbon contributes efficiently to additional molecular opacity sources, primarily C$_2$, CN, and even C$_2$H$_2$ at cooler temperatures. As a consequence, the atmospheric structure, radius and thus the effective temperature of the merger remnant are expected to depend sensitively on the carbon excess. However, once the atmosphere becomes strongly carbon dominated (C/O $\gg$ 1), the sensitivity of the atmospheric opacity to further increases in C/O is expected to weaken. \citet{Mehla2025} have observed wide variations of C/O with values ranging between 10 and 80. While C/O controls the dominant atmospheric opacity, metallicity is also expected to influence the atmospheric structure.

In this study, we searched for dLHdC stars within the same temperature and luminosity ranges as those observed in the Milky Way. Accordingly, we focused our analysis on the BL subsample defined by \citet[see their Fig.~2]{Luri2021}. Therefore, it is possible that our selection misses part of the dLHdC population by restricting the search to a specific region of the Hertzsprung--Russell diagram. The Galactic dLHdC star A166 -- which we did not use as a reference dLHdC in our analyses owing to its unknown distance -- is the only such example with its cool effective temperature (as indicated by its $(G_{BP} - G_{RP})_0\sim$1.8 mag colour) and low brightness that positioned it outside the BL area in an HR diagram \citep[see their Fig.10]{Tisserand2022}. However, \citet[][see their Sect.~4.6]{Tisserand2022} reported a low level of dust production for A166, together with additional indicators --- notably its $^{16}$O/$^{18}$O ratio \citep{Karambelkar2022} and long-timescale photometric variability --- more consistent with the RCB subclass, casting some doubt on its nature: it may instead be an RCB star observed during a quiescent phase. If so, no confirmed dLHdC star is currently known outside our selection region, further arguing against a substantial hidden population of cool, faint dLHdC stars.

Considering a direct effect of metallicity on the effective temperature distribution, \citet{Crawford2020} pointed out that higher-metallicity HdC atmospheres may be expected to reach cooler effective temperatures. Consequently, metal-rich dLHdC stars may preferentially overlap with the cooler and more populated red giant branch region, whereas metal-poor dLHdC stars are expected to appear at higher effective temperatures. 

However, the known Galactic dLHdC stars have been identified across all Galactic substructures \citep{Tisserand2024a}, including the halo, bulge, and thin and thick disks, suggesting that they already span a broad range of metallicities. This is potentially also reflected in their observed wide colour distribution. Since the Magellanic Clouds are, on average, more metal-poor than the Milky Way, one may instead expect their dLHdC population to be shifted towards bluer colours. Such a shift should nevertheless remain largely encompassed within the BL region considered in this study, arguing against a major selection bias arising solely from metallicity-induced temperature shifts.

\subsection{The dLHdC-to-RCB ratio across Galactic substructures}
\label{sec:substructures}

Since metallicity is central to our interpretation, it is natural to ask
whether the relative incidence of the two HdC subclasses already varies
among the Galactic stellar populations. Using the kinematic classification
of \citet{Tisserand2024a}, the numbers of RCB and dLHdC stars assigned to
each substructure are listed in Table~\ref{tab:substructures}.

Two effects make this comparison difficult. First, the Galactic censuses of
the two subclasses do not have the same effective depth: dLHdC stars are
intrinsically fainter than RCB stars, so the dLHdC search probed a smaller
volume, and it did not cover the most highly reddened regions close to the
Galactic plane and within the bulge\footnote{The search of Tisserand et al. (2022) was restricted to lines of sight with E(B-V) $<$ 1 mag.}. Second, while a reliable distance estimate could be obtained for nearly all known Galactic dLHdC stars, this
is not the case for a majority of RCB stars: their Gaia DR3 astrometric
solutions are degraded by the PSF chromaticity effect induced during their
photometric declines, so that their parallaxes, and hence their distances,
are incorrect \citep{Tisserand2024a}. Both censuses are therefore affected by selection effects, and the ratios listed in Table~\ref{tab:substructures} cannot be read directly as relative incidences.

The halo is the substructure least affected by the second of these two
biases. Among the seven halo RCB stars, six (all but U Aqr) had astrometric
fits corrupted by the PSF chromaticity effect. Nonetheless, they could be
assigned a halo membership from their photometry and a plausible distance
estimate, so that the halo RCB census is not strongly depleted by the
astrometric selection.

On the dLHdC side, only one of the four objects classified in the halo is a
pure dLHdC star: the three others show a low but non-zero level of dust
production, and may instead be RCB stars observed between two active
phases, as already discussed for A166 in Sect.~\ref{sec:discus_1}. The
remaining object, HE 1015-2050, was identified as a hydrogen-deficient
carbon star in a deep spectroscopic survey of faint high-latitude carbon
stars \citep{Goswami2010}, and lies at a distance in excess of 50 kpc, in
the far outer regions of the Galaxy \citep{Tisserand2024a}. Its dLHdC
classification was discussed in \citet{Tisserand2022}.

Taken at face value, the halo thus presents about one dLHdC star for every
seven RCB stars, markedly below the near-unity ratio measured for the
Galactic population as a whole, and in the same direction as the deficit
reported here for the Magellanic Clouds. This is the behaviour expected if,
as we argue in Sect.~\ref{sec:discus_popsynth}, lower metallicity biases
the total WD merger mass distribution towards higher values, which in turn
favours HdC stars of higher luminosity. The two dedicated surveys, however,
do not share the same sky coverage: the four northern halo RCB stars
AO Her, NSV 11154, Z UMi and WISE J004822.34+741757.4 have no counterpart
in the dLHdC census, since \citet{Tisserand2022} searched only the southern
sky and no equivalent northern search for dLHdC stars has yet been carried
out.

The numbers involved nevertheless remain small  and the halo ratio
further depends on whether the three low-dust-production objects are
counted as dLHdC stars, which would raise it to roughly 1:2. A larger and
homogeneously selected sample of halo HdC stars, together with atmospheric
metallicity measurements, will be required to establish whether the subclass
ratio varies systematically with metallicity within the Galaxy.

\begin{table}
\caption{Numbers of Galactic RCB and dLHdC stars classified within each
Galactic substructure by \citet{Tisserand2024a}. Objects showing a low
level of dust production, whose dLHdC nature is uncertain, are given in
parentheses.}
\label{tab:substructures}
\centering
\begin{tabular}{lcc}
\hline\hline
Substructure & RCB & dLHdC \\
\hline
Thin disk  & 7 & 5 (+1) \\
Thick disk & 7 & 6 \\
Bulge      & 9 & 16 \\
Halo       & 7 & 1 (+3) \\
\hline
\end{tabular}
\end{table}

\subsection{Population synthesis effects \label{sec:discus_popsynth}}

Differences in the underlying stellar populations may also affect the relative production rates of dLHdC and RCB stars. In the double-degenerate scenario, population synthesis simulations \citep{Karakas2015, Tisserand2022, Tisserand2024a} indicate that the dominant and more massive merger channel (hereafter HybCO) involves the coalescence of a hybrid COHe WD \citep{Tutukov1996, Staff2018, Zenati2019, Shiber2024} with a CO WD. This HybCO merger channel is characterised by a total mass distribution skewed towards higher masses, peaking at $\sim$0.95~M$_\odot$, extending up to $\sim$1.05~M$_\odot$ and with a tail extending down to $\sim$0.7~M$_\odot$, whereas other merger channels show a narrower distribution centred on $\sim$0.65~M$_\odot$ \citep[see their Fig.~13]{Tisserand2022}. These main distribution features persist both at solar metallicity (Z$_\odot$=0.02 in the adopted \textit{StarTrack} model grid; \citealt{Belczynski2008, Ruiter2014}) and at lower metallicity representative of the SMC (Z$=$0.004).

Interestingly, the HybCO merger channel is also the only one exhibiting a bimodal delay time distribution (DTD), with a prominent peak around 1~Gyr (spanning $\sim$0.5--2~Gyr) and a broader component extending from $\sim$5~Gyr up to a Hubble time. Population synthesis models further predict that, at solar metallicity, about 60\% of these mergers occur at short delay times, increasing to $\sim$70\% at Z$=$0.004 \citep[see their Fig.~19 for Z$=$0.02]{Tisserand2024a}.

Considering these results together with the star formation history of the Milky Way, and the fact that RCB and dLHdC stars are observed across all Galactic components, the HybCO merger channel emerges from the population synthesis, without any tuning, as the most plausible dominant formation pathway for both HdC subclasses. The bimodality of its delay time distribution is required by the presence of HdC stars in the oldest Galactic substructures, the bulge and the halo \citep{Tisserand2024a}: the other channels produce mergers only at delay times of order 1 Gyr and would confine HdC stars largely to the thin disk and the bulge. Its broad total mass range ($\sim$0.7--1.05~M$_\odot$) is likewise required by the wide range of absolute magnitudes observed among HdC stars, since the luminosity of the merger remnant is closely tied to the total mass \citep{Crawford2024}, whereas the narrow distributions of the other channels, peaking near 0.65~M$_\odot$, could not reproduce it. That range is moreover consistent with the only direct mass estimates available for any HdC star: the pulsation modelling of \citet{Saio2008} yields masses of $\sim$0.8--0.9~M$_\odot$ for RCB stars, above the peak characteristic of the other channels, although such determinations exist only for the RCB stars with well-characterised pulsations and none is available for dLHdC stars. Finally, HybCO is by far the most numerous channel. We stress nevertheless that no existing calculation demonstrates that it dominates the formation of the RCB and dLHdC subclasses individually, and that a contribution of the lower-mass channels to the dLHdC population cannot be excluded.

The Magellanic Clouds, however, exhibit different star formation histories. In the LMC, \citet[see their Figs.~5 and ~17]{Mazzi2021} identified two relatively recent episodes of enhanced star formation occurring at $\sim$1.8~Gyr and $\sim$0.8~Gyr ago, both predominantly located within the bar region. An earlier major episode around 10~Gyr ago produced a more spatially extended stellar population, similar to the distribution of RR~Lyrae stars \citep[see their Fig.~B2 or Fig.~5, panel i16]{Mazzi2021}.

By convolving the HybCO merger DTD with the LMC star formation history from \citet{Mazzi2021} (integrated over all VMC survey, see their Fig.~17), we estimate that about 80\% of HdC stars in the LMC should originate from the short delay time merger component, i.e., the younger stellar population. This interpretation is qualitatively consistent with the observed spatial distribution of RCB stars in the LMC, most of which (all but four) are concentrated along the bar \citep[see their Fig.~9]{Tisserand2009}.

Within this framework, the observed scarcity of dLHdC stars in the Magellanic Clouds may indicate that short delay time HybCO mergers preferentially produce higher total WD masses at low metallicity, thereby favouring the formation of RCB stars during the HdC phase. In support of this interpretation, Magellanic RCB stars are observed to be systematically brighter than their Galactic counterparts \citep[see their Fig.~10]{Tisserand2022}, which is consistent with a higher total WD mass distribution \citep{Crawford2024}.

\begin{figure}[h]
    \centering
    \includegraphics[width=1\linewidth]{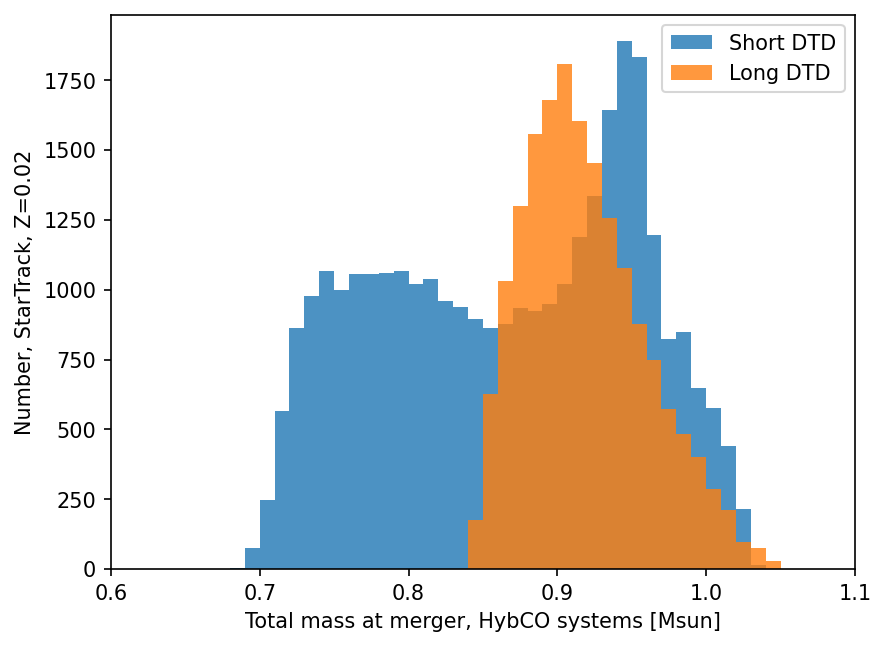}
    \includegraphics[width=1\linewidth]{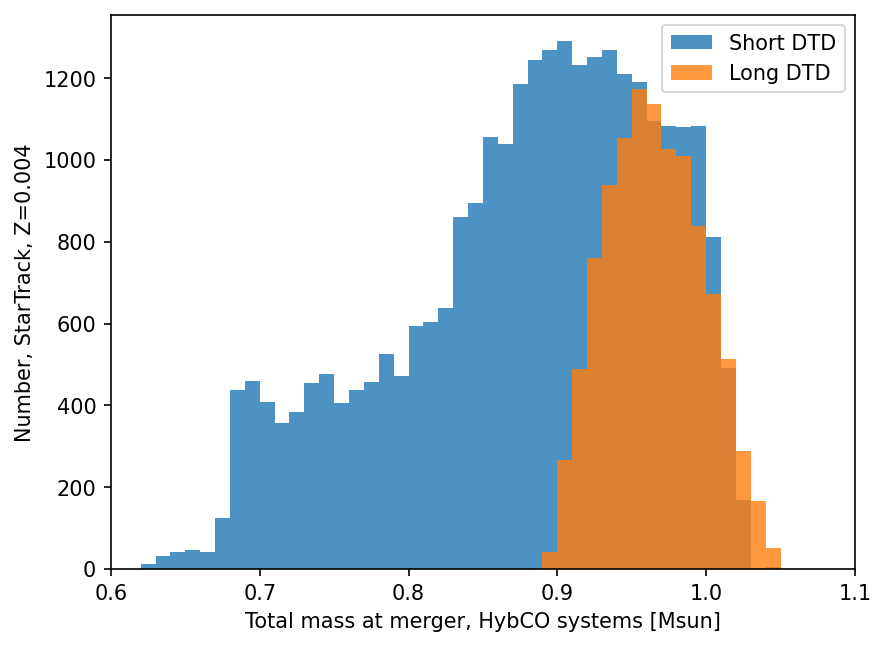}
    \caption{Total mass distributions of HybCO WD mergers for the short (blue) and long (orange) delay time regimes. Results from \textit{StarTrack} population synthesis simulations performed at low metallicity (Z$=$0.004, bottom) and solar metallicity (Z$=$0.02, top). }
    \label{fig:TotalMassWD_DTD}
\end{figure}

From an evolutionary perspective, short delay time HybCO mergers are expected to evolve from stellar systems undergoing a common-envelope phase while the mass-losing secondary is on the red giant branch, rather than on the asymptotic giant branch as in longer delay time systems \citep{Karakas2015, Tisserand2022, Tisserand2024a}. In the short delay time regime, the secondaries are more massive ($\sim$1.5--3.5~M$_\odot$ at ZAMS, i.e., the zero-age main sequence, Fig.~\ref{fig:MassDistrib_ZAMS_Annexe}) than those of the long delay time regime, since the delay time is dominated by the nuclear lifetime of the secondary. The resulting CO WD masses span a broad range ($\sim$0.40--0.70~M$_\odot$, Fig.~\ref{fig:MassDistrib_2WD_Annexe}), making this the only component extending into the low-mass regime associated with dLHdC stars, whereas in the long delay time regime, the lower-mass secondaries ($\sim$1.2--2.0~M$_\odot$) reach the AGB with well-developed cores before the common-envelope phase, producing a narrow CO WD mass distribution confined to $\sim$0.55--0.70~M$_\odot$. Since the mass of the hybrid WD primary is nearly constant ($\sim$0.32--0.35~M$_\odot$, Fig.~\ref{fig:MassDistrib_2WD_Annexe}) across all delay times and metallicities, the total merger mass distribution directly reflects this contrast in secondary masses. Metallicity then acts on top of this structure: metal-poor progenitors produce more massive WDs for a given initial mass \citep{Romero2015}, with differences of up to $\sim$0.1~M$_\odot$, shifting the short delay time mass distribution towards higher masses at Z=0.004 (Fig.~\ref{fig:TotalMassWD_DTD}) and depleting the low-mass tail from which dLHdC stars would otherwise originate.

The resulting total mass distributions of HybCO mergers, shown separately for the short and long delay time components at Z=0.02 and Z=0.004, are presented in Fig.~\ref{fig:TotalMassWD_DTD}. At low metallicity, the short delay time component is strongly skewed towards higher total masses. Since the luminosity of the post-merger supergiant increases with the total mass of the merged system \citep{Crawford2024}, and since dLHdC stars are about two magnitudes fainter than RCB stars \citep{Tisserand2022}, such a skew is expected to favour the production of luminous RCB stars at the expense of the fainter dLHdC population. The underlying mass distributions of the individual WDs and of their ZAMS progenitors, for both metallicities and both delay time components, are detailed in Appendix~\ref{appendix:MassDistrib}.

This scenario provides a plausible explanation for the deficit of dLHdC stars observed in the LMC, highlighting the role of stellar population effects and metallicity-dependent binary evolution in shaping the observed HdC subclasses. The SMC, whose star formation history differs markedly from that of the LMC, offers a differential test of this framework.

In the SMC, contrary to the LMC, the most prominent star formation episode occurred between 4.5 and 6~Gyr ago, peaking at a level roughly twice that of the star formation rate (SFR) from $\sim$1~Gyr ago \citep{Rubele2018}. By convolving the HybCO merger DTD with the SMC star formation history from \citet{Rubele2018} (the global SFR integrated over all VMC survey, see their Fig.~11), we estimate that only about two-thirds of the SMC HdC stars should originate from the short delay time merger component, and one-third from the long one. Unlike in the LMC, where the recent star formation bursts coincide with the short delay time peak of the HybCO DTD and amplify it, the SMC's dominant star formation episode at $\sim$4.5--6~Gyr feeds the long delay time component, leaving the present-day merger mix close to the intrinsic proportions of the DTD. The short delay time channel therefore still dominates --- again favouring the higher-mass regime of HdC formation, and thus a low dLHdC-to-RCB number ratio --- though to a markedly lesser extent than in the LMC.

We stress that this convolution predicts the relative contributions of the short- and long-delay merger components, not the dLHdC-to-RCB number ratio itself. The latter depends in addition on how merger parameters map onto the two subclasses, and on the relative durations of the two phases, which enter explicitly through Eq.~\ref{eq:lifetime} (Sect.~\ref{sec:lifetimes}). Our calculation therefore indicates the direction in which the subclass ratio is expected to shift, and the ordering expected between the two Clouds, but it does not by itself provide a quantitative prediction of the observed ratio.

This framework does, however, make a differential prediction: the dLHdC-to-RCB ratio should be higher in the SMC than in the LMC. The observed ratios, 1:24 in the LMC and 1:6 in the SMC, indeed follow the predicted ordering. However, with only one dLHdC star identified in each Cloud and only six RCB stars known in the SMC, the two ratios remain statistically indistinguishable, and the small-number statistics make any quantitative interpretation tentative.

Interestingly, the newly discovered SMC dLHdC star, SMSS0055-72, exhibits several characteristics consistent with an origin in the long delay time merger component. In particular, its Sr-rich spectrum suggests an association with an old, metal-poor stellar population (see Sect.~\ref{sec:SMSS0055}). If this interpretation is correct, the progenitor stellar binary system would have formed at least $\sim$5~Gyr ago --- and possibly as early as $\sim$10~Gyr ago --- with the final HybCO white-dwarf merger occurring only recently. SMSS0055-72 would therefore be a member of the long delay time merger population.

This interpretation, however, sits in tension with our lowest-metallicity model grid: at Z=0.004 the long delay time component is confined to the high-mass regime ($\sim$0.9--1.05~M$_\odot$, peaking near 0.96~M$_\odot$, see Fig.~\ref{fig:TotalMassWD_DTD}), where RCB rather than dLHdC stars are expected. Two avenues could reconcile this. First, if SMSS0055-72 is indeed a long delay time, low-mass merger remnant, its existence would suggest that the long delay time mass distribution at metallicities below Z=0.004 — consistent with the old, metal-poor stellar population of the SMC — is less sharply confined at the low-mass end than our Z=0.004 grid indicates. Population synthesis at metallicities below Z=0.004 would be required to test this. Second, and alternatively, the association between the Sr-rich signature and an old, metal-poor population --- established empirically from the Galactic halo location of most known Sr-rich HdC stars \citep{Crawford2022} --- may not be causal: the s-process enrichment could instead reflect the specific nucleosynthetic conditions of the merger itself rather than the age of the progenitor system. In that case, SMSS0055-72 does not need to descend from the long delay time component, and no tension with the mass distributions arises. Distinguishing these possibilities will require both lower-metallicity population synthesis and a detailed abundance analysis of SMSS0055-72.

\subsection{Relative lifetimes of the dLHdC and RCB phases  \label{sec:lifetimes}}

We note that the observed number ratio depends not only on the relative birthrates of low- and high-mass merger remnants, but also on the relative durations of the dLHdC and RCB phases. From their post-merger evolution simulations at low metallicity (10\% solar),
\citet{Crawford2024} found that dLHdC stars are expected to have shorter lifetimes than RCB stars ($0.5$--$1.2\times10^{5}$~yr versus $1$--$5\times10^{5}$~yr), with the duration of the HdC phase decreasing with decreasing total WD mass and/or increasing WD mass ratio. The two subclasses also evolve differently: RCB stars remain within their
luminosity and effective temperature range for most of the HdC phase before evolving rapidly bluewards, whereas dLHdC stars evolve continuously towards warmer temperatures.

In a steady-state regime, the number of stars observed in each subclass scales with the product of its formation rate and the duration of the corresponding phase. The observed number ratio can therefore be expressed as:

\begin{equation}
\frac{N_{\mathrm{dLHdC}}}{N_{\mathrm{RCB}}} =
\frac{R_{\mathrm{dLHdC}}}{R_{\mathrm{RCB}}} \times
\frac{\tau_{\mathrm{dLHdC}}}{\tau_{\mathrm{RCB}}},
\label{eq:lifetime}
\end{equation}

where $R$ denotes the birthrate of merger remnants entering each subclass and $\tau$ the duration of the corresponding HdC phase. The lifetimes calculated by \citet{Crawford2024} correspond to $\tau_{\mathrm{dLHdC}}/\tau_{\mathrm{RCB}} \approx 0.25$--$0.5$, so that the observed dLHdC-to-RCB ratio is expected to be a factor of
$\sim$2--4 lower than the underlying ratio of merger birthrates. However, this suppression only contributes to the \emph{differential} deficit between the Milky Way and the Magellanic Clouds if the lifetime asymmetry is itself metallicity-dependent --- if it is of similar amplitude at solar metallicity, it affects the Galactic and Magellanic observed ratios equally and cancels out of the comparison. This dependence remains unconstrained by the currently available single-metallicity simulations. If the lifetime asymmetry were weaker at solar metallicity, it would act in the same direction as the merger-mass effect discussed above and further reduce the expected
Magellanic dLHdC-to-RCB ratio.

\section{Conclusion \label{sec:concl}}

We have conducted the first systematic search for dustless hydrogen-deficient carbon stars in the Magellanic Clouds, combining classical colour--magnitude selection with a random-forest classifier, both anchored to the known Galactic dLHdC population, and following up all selected candidates spectroscopically with WiFeS on the ANU 2.3m telescope. The overall detection efficiency of the search is estimated at $\sim$96\% (Sect.~\ref{sec:detecteff}). The two analyses proved complementary: we discovered the first two Magellanic dLHdC stars, SP77~29-5 in the LMC --- identified by both methods --- and SMSS~005551.43-723139.2 in the SMC, recovered only by the random-forest analysis. SP77~29-5 is of spectral type HdC3 C$_2$ 4.5, closely resembling the Galactic dLHdC star B563, while SMSS0055-72 is warmer ($\sim$6700~K) and belongs to the rare Sr-rich subclass,  predominantly found in the Galactic halo and the LMC outskirts.

These discoveries establish a dLHdC-to-RCB number ratio of $\sim$1:15 in the Magellanic Clouds (1:24 in the LMC, 1:6 in the SMC), in sharp contrast with the near-unity ratio measured in the Milky Way. Given our detection efficiency of 96\%, this deficit cannot be attributed to observational incompleteness for dLHdC stars with properties similar to the known Galactic population. Two regimes nevertheless remain less constrained: cool, faint dLHdC stars analogous to A166, lying outside the blue-loop selection region, and the cool, nitrogen-poor dLHdC stars whose colours, according to our synthetic grid, would fall outside our selection.

We propose instead that the deficit reflects the interplay between the star formation histories of the Clouds and the metallicity-dependent mass distribution of white dwarf mergers. By convolving the delay time distribution of the dominant HybCO merger channel with each galaxy's star formation history, we estimate that $\sim$80\% of LMC HdC stars and about two-thirds of SMC HdC stars originate from short delay time mergers. Population synthesis simulations show that at low metallicity these short-delay mergers are strongly biased towards higher total WD masses --- the regime expected to produce the more luminous, dusty RCB stars --- at the expense of the lower-mass regime associated with dLHdC stars. The shorter predicted lifetimes of the dLHdC phase \citep{Crawford2024} may further suppress the observed ratio, if this asymmetry strengthens at low metallicity. This picture is supported by the systematically higher luminosities of Magellanic RCB stars compared to their Galactic counterparts, and it makes a differential prediction --- a higher dLHdC-to-RCB ratio in the SMC than in the LMC --- that the observed ratios indeed follow, although the current small-number statistics prevent a significant test. 

We stress that this calculation predicts the relative contributions of the two delay time components, and thus the direction in which the dLHdC-to-RCB ratio is expected to shift, rather than the value of the ratio itself, which depends in addition on the mapping between merger parameters and HdC subclasses and on the relative durations of the dLHdC and RCB phases.

The peculiar chemistry of SMSS0055-72 --- in particular its Sr-rich spectrum --- makes it a valuable test case for this framework, though not yet a decisive one: whether its s-process signature traces an old, metal-poor progenitor population ($>$5 Gyr), and thus a long delay time merger, or instead reflects the nucleosynthetic conditions of the merger itself, remains open (Sect.~\ref{sec:discus_popsynth}). Resolving this, together with the extension of population synthesis to metallicities below Z=0.004, is needed to establish how individual dLHdC stars map onto the merger mass and delay time distributions.

The ratio of dLHdC to RCB stars thus emerges as a sensitive probe of both the delay time distribution of white dwarf mergers and the metallicity dependence of binary evolution. High-resolution spectroscopic follow-up of the two new Magellanic dLHdC stars --- in particular abundance analyses and $^{16}$O/$^{18}$O isotopic ratios --- would provide a direct test of their merger origin in a low-metallicity environment.

On the observational side, two recent photometric data sets open a clear path to deeper and purer searches. The fourth data release of the S-PLUS survey \citep{Herpich2024_SPLUS} covers the Magellanic Clouds with PSF photometry in the 12-band Javalambre filter system, whose narrow filters directly sample the key HdC diagnostics --- the Balmer jump region (with J0378, J0395, J0410 and classic u filter), the H$\alpha$ line (J0660), and the C$_2$ Swan (0,0) bandhead (J0515) --- effectively providing very-low-resolution spectrophotometry for every star and promising a substantial reduction of the variable-star contamination that dominated our candidate samples. In parallel, the deep DECam $ugriz$ photometry of the SMASH survey \citep{Nidever2017_SMASH} reaches several magnitudes below both our faint-magnitude limit and the SkyMapper $u$-band sensitivity that nearly excluded SMSS0055-72 from our selection. Together, these data sets will allow the search to be extended to the cooler, fainter dLHdC regime discussed in Sect.~\ref{sec:discus_1}, and to test whether the dearth reported here persists at the low-luminosity end of the HdC population.

On the theoretical side, population synthesis calculations coupling metallicity-dependent binary evolution with post-merger stellar evolution models will be required to predict the HdC subclass ratio — and the relative durations of the dusty and dustless phases — quantitatively across different galactic environments. We would also encourage independent population synthesis studies using other codes and other sets of assumptions, in particular for the common-envelope efficiency, to test how robust the predicted merger mass and delay time distributions are to these choices.

\begin{acknowledgements}
AR's work was supported by the ARC grant number FT170100243. AM is supported by the ARC Discovery Early Career Research Award (DE230100055). The research was based on data acquired at the ANU 2.3-metre telescope. The automation of the telescope was made possible through an initial grant provided by the Centre of Gravitational Astrophysics and the Research School of Astronomy and Astrophysics at the Australian National University and through a grant provided by the Australian Research Council through LE230100063. The Lens proposal system is maintained by the AAO Research Data \& Software team as part of the Data Central Science Platform. We acknowledge the traditional custodians of the land on which the telescope stands, the Gamilaraay people, and pay our respects to elders past and present. This work presents results from the European Space Agency (ESA) space mission Gaia. Gaia data are being processed by the Gaia Data Processing and Analysis Consortium (DPAC). Funding for the DPAC is provided by national institutions, in particular the institutions participating in the Gaia MultiLateral Agreement (MLA). This research has made use of the SIMBAD database, operated at CDS, Strasbourg, France, of TOPCAT (Taylor, M., 2005, ASPC, 347, 29), an interactive graphical viewer and editor for tabular data, and of the Spanish Virtual Observatory (\url{https://svo.cab.inta-csic.es}) project funded by MCIN/AEI/10.13039/501100011033 through grant PID2023-146210NB-I00.

\end{acknowledgements}

\bibliographystyle{aa}
\bibliography{RCB_Citations}


\appendix

\section{HybCO merger mass distributions from population synthesis \label{appendix:MassDistrib}}

This appendix presents the detailed mass distributions underlying the HybCO merger channel discussed in Sect.~\ref{sec:discus_popsynth}, obtained from the \textit{StarTrack} population synthesis simulations \citep{Belczynski2008, Ruiter2014, Ruiter2019} at solar metallicity (Z=0.02) and at a metallicity representative of the SMC (Z=0.004). All distributions are shown separately for the short ($\sim$0.5--2~Gyr) and long ($\gtrsim$5~Gyr) delay time components of the DTD.

Our StarTrack binary population synthesis model data were the same as those used in \citet{Tisserand2022} and \citet{Tisserand2024a}. We note that our stellar populations are `simple', meaning that there is no recycling of elements taken into account, thus no subsequent (enriched) generation of stars: all stars in a simulation are born with the same metallicity. Two simple stellar populations were examined in this study: one with approximately solar (Z=0.02) metallicity, while the other has sub-solar (Z=0.004) metallicity, which is more appropriate for metal-poor stellar populations such as the Magellanic Clouds.  The ZAMS mass distributions of the binary stars were selected such that the initially more massive star mass is drawn from a broken three-component power-law in the range $0.8 < M_1 < 150~M_\odot$, see \citet{Ruiter2019} for general details. To select the secondary mass, we choose the mass ratio from a mass ratio distribution guided by \citet{Sana2012} with the secondary mass falling in the range $0.5 < M_2 < 150~M_\odot$, and $M_2 < M_1$. Initial distributions observe the following forms: $q$ is drawn from a flat distribution with $0.1 < q < 1$, orbital period $\propto (\log P)^{-0.55}$ and the initial eccentricity distribution follows $\propto e^{-0.42}$, where $e$ is the eccentricity.

While orbital properties as described in \citet{Sana2012} have been observed to work relatively well for massive stars, it is worthwhile exploring other types of initial binary parameter distributions in future studies as applied to the intermediate-mass stars studied here. In terms of treatment of common envelope events, we adopt a `standard' (constant $\alpha \times \lambda$) prescription utilising the energy balance formalism of \citet{Webbink1984} \citep[see][Sect. 2 for more details]{Ruiter2019}. Since in our simulations we are most concerned with stars of intermediate mass, we adopt a slightly higher efficiency than the `old standard' value of $\alpha \times \lambda = 1$, and instead use $\lambda = 2$ and $\alpha = 1$, based on enthalpy arguments \citep{Ivanova2011}. For a general description of how mass transfer is treated in binary population synthesis codes, in particular for stars that are of low- to intermediate-mass, the interested reader is encouraged to see the population synthesis code comparison study by \citet{Toonen2014}, while more comprehensive details on how mass transfer is handled in the StarTrack code can be found in \citet{Belczynski2008}.

\begin{figure}[ht]
    \centering
    \includegraphics[width=0.495\linewidth]{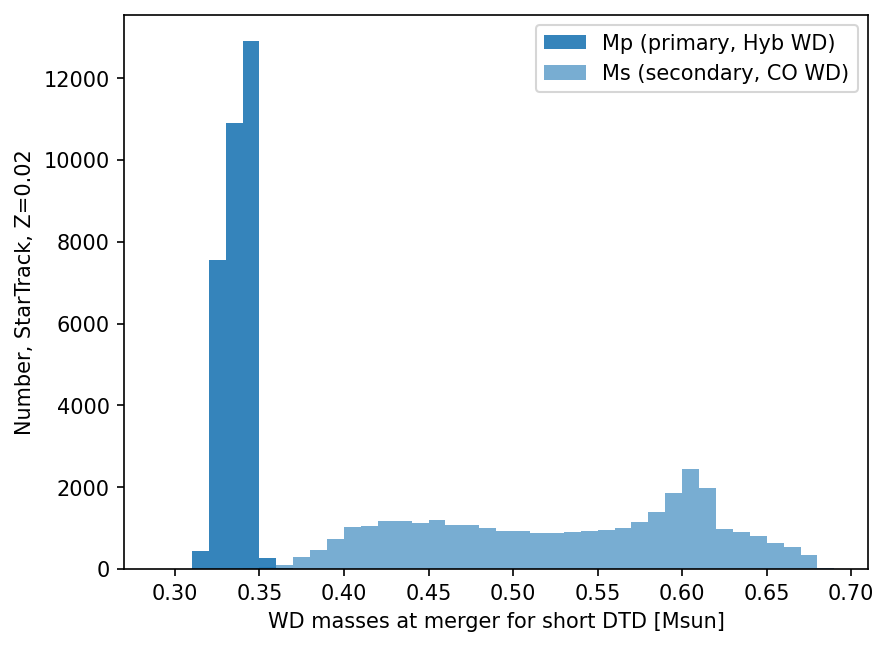}
    \includegraphics[width=0.495\linewidth]{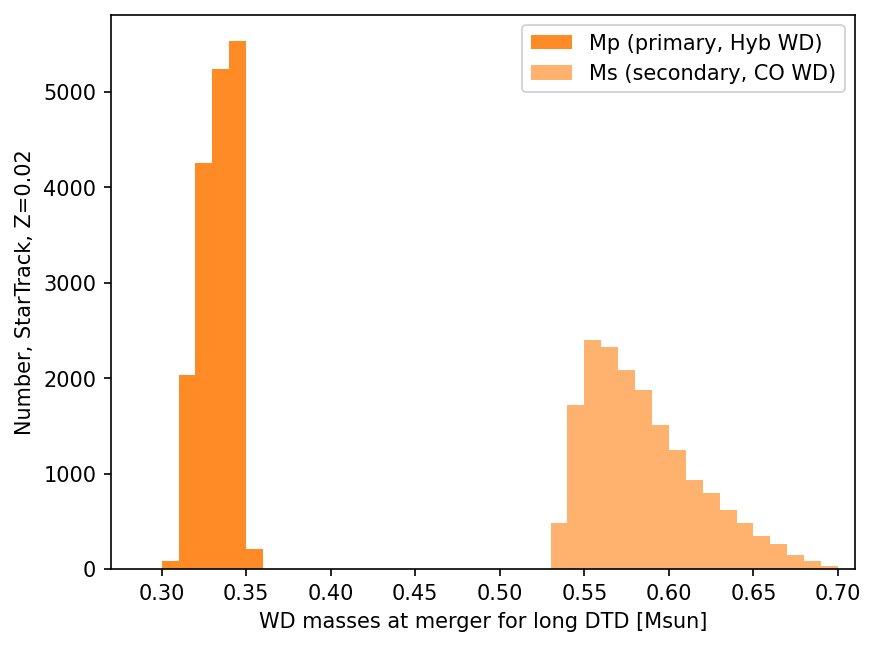}
    \includegraphics[width=0.495\linewidth]{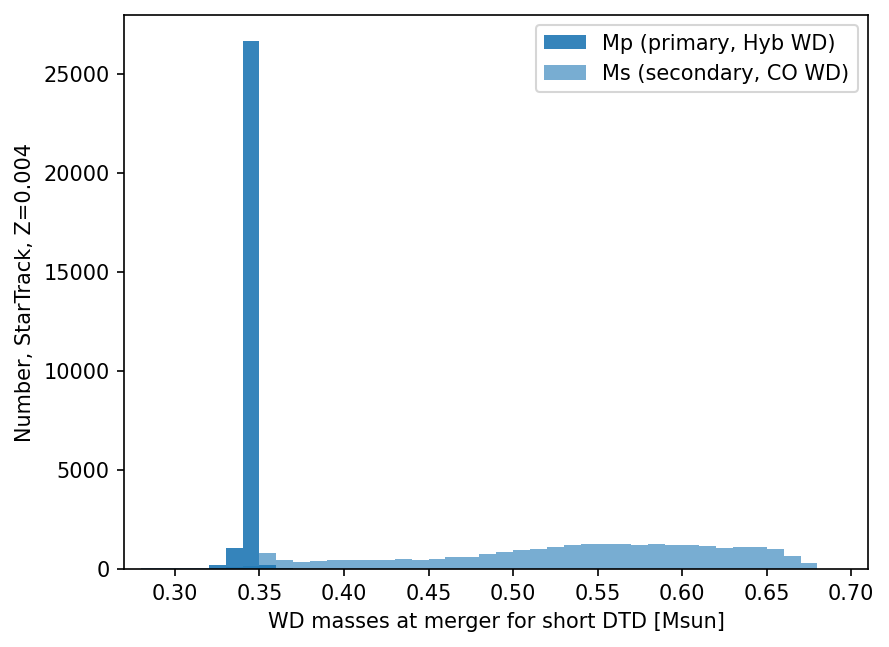}
    \includegraphics[width=0.495\linewidth]{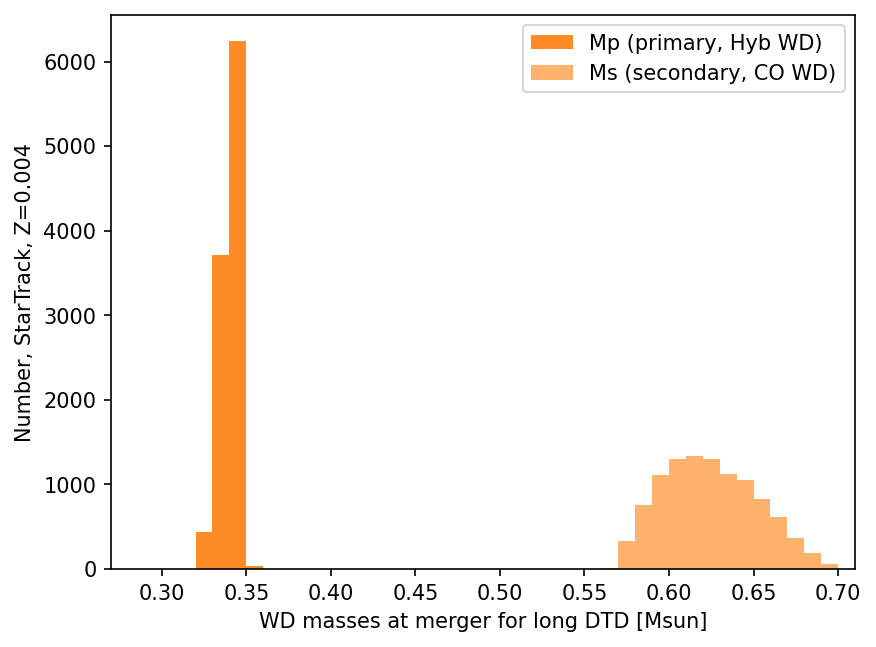}
    \caption{Mass distributions of the two WDs involved in HybCO mergers, for the short (left) and long (right) delay time regimes. Results from \textit{StarTrack} population synthesis simulations made at low metallicity (Z$=$0.004, bottom) and solar metallicity (Z$=$0.02, top).}
    \label{fig:MassDistrib_2WD_Annexe}
\end{figure}

Figure~\ref{fig:MassDistrib_2WD_Annexe} shows the individual masses of the two white dwarfs at the time of the merger. The mass of the primary, the hybrid COHe WD, is strongly peaked near $\sim$0.32--0.35~M$_{\odot}$ in all four combinations of metallicity and delay time: it is a structural property of the hybrid WD formation channel and is largely insensitive to both parameters. The variation in total merger mass is instead carried almost entirely by the secondary CO~WD. In the short delay time regime, its mass spans a broad range ($\sim$0.40--0.70~M$_{\odot}$), whereas in the long delay time regime it is confined to a narrow interval near $\sim$0.55--0.70~M$_{\odot}$. In both cases, the secondary CO~WD mass distribution is more concentrated towards higher values at Z=0.004. The metallicity dependence of the total merger mass distribution presented in Fig.~\ref{fig:TotalMassWD_DTD} therefore reflects the behaviour of the secondary CO~WD mass.

\begin{figure}[ht]
    \centering
    \includegraphics[width=0.495\linewidth]{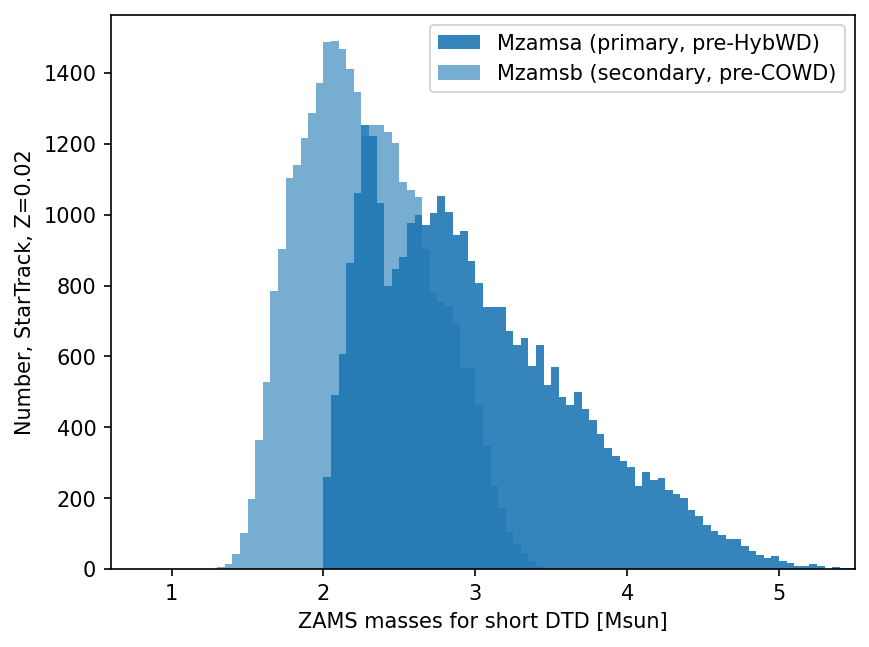}
    \includegraphics[width=0.495\linewidth]{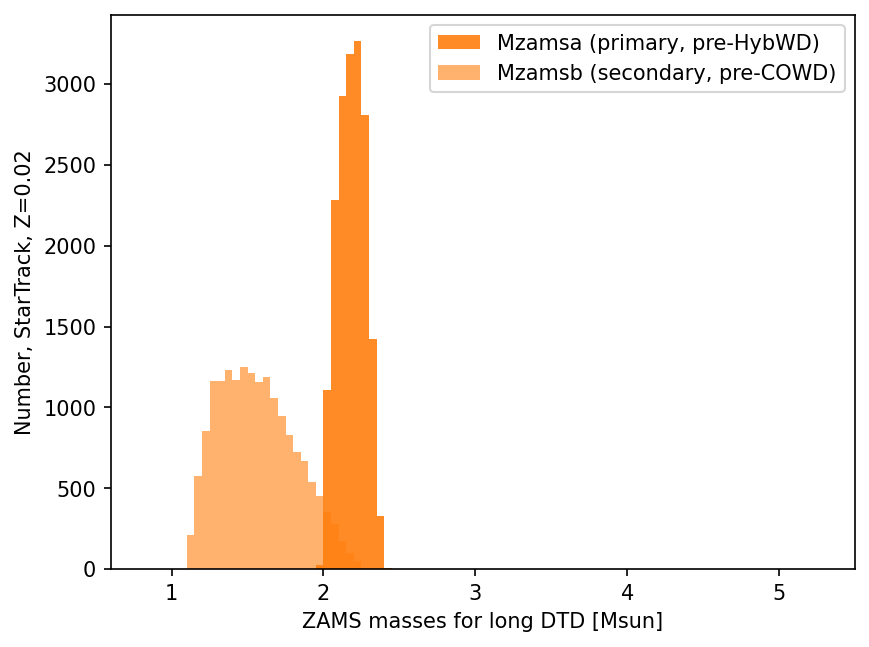}
    \includegraphics[width=0.495\linewidth]{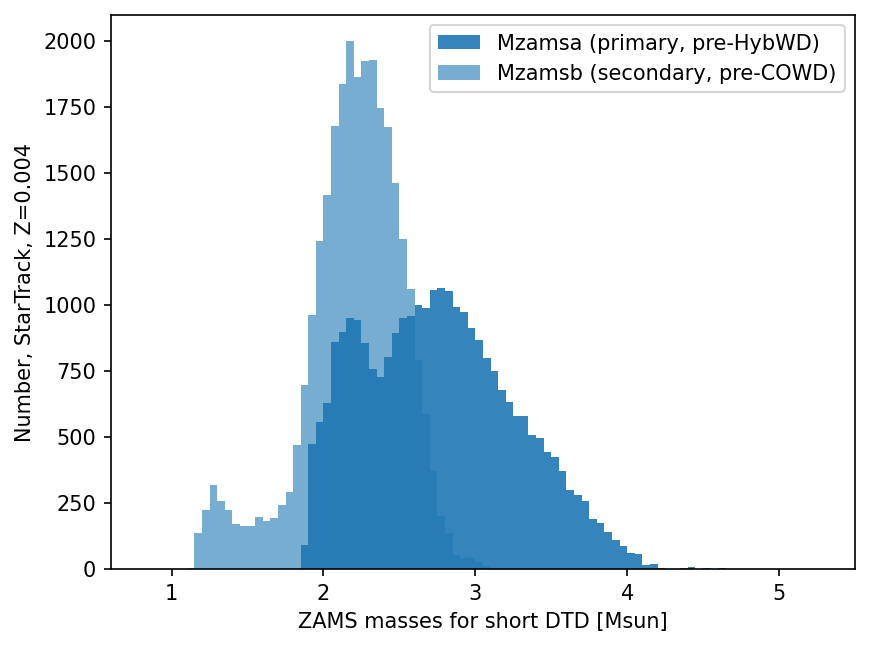}
    \includegraphics[width=0.495\linewidth]{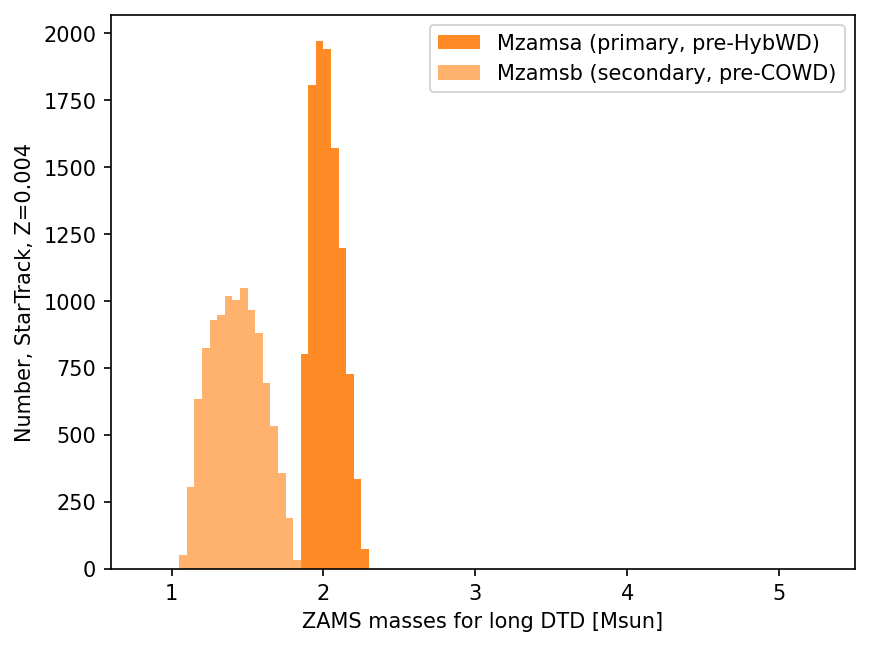}
    \caption{Mass distributions of the original zero-age main sequence stellar components resulting in HybCO WD mergers, for the short (left) and long (right) delay time regimes. Results from \textit{StarTrack} population synthesis simulations made at low metallicity (Z$=$0.004, bottom) and solar metallicity (Z$=$0.02, top).}
    \label{fig:MassDistrib_ZAMS_Annexe}
\end{figure}

Figure~\ref{fig:MassDistrib_ZAMS_Annexe} presents the corresponding zero-age main sequence (ZAMS) masses of the two components of the progenitor binaries. Short delay time mergers originate from comparatively massive systems, with primary masses spanning $\sim$2.0--5.0~M$_{\odot}$ and secondary masses $\sim$1.5--3.5~M$_{\odot}$, whereas long delay time mergers descend from a much narrower progenitor population, with primaries confined near $\sim$2.0--2.3~M$_{\odot}$ and secondaries of only $\sim$1.2--2.0~M$_{\odot}$. This dichotomy directly reflects the origin of the delay time itself: it is dominated by the nuclear lifetime of the secondary, so that long delay times require low-mass secondaries, which in turn produce the narrow, well-defined CO~WD mass distribution seen in Fig.~\ref{fig:MassDistrib_2WD_Annexe}. At lower metallicity, both ZAMS distributions shift modestly, consistent with the metallicity dependence of the initial-to-final mass relation \citep{Romero2015} invoked in Sect.~\ref{sec:discus_popsynth}.

\end{document}